\documentclass[twocolumn,aps,prl,showpacs,amsmath,superscriptaddress,longbibliography,notitlepage]{revtex4-1}
\usepackage{amssymb}
\usepackage{mathrsfs}
\usepackage{graphicx}
\usepackage{float}
\usepackage[caption=false]{subfig}
\usepackage{epstopdf}
\usepackage[normalem]{ulem}
\usepackage{verbatim}
\usepackage{color}
\usepackage{bm}

\definecolor{orange}{RGB}{255, 100, 0}
\def\CH{\textcolor{black}}

\def\GJ{\textcolor{black}}
\def\blue{\textcolor{blue}}
\def\red{\textcolor{red}}

\begin{document}

\newcommand{\norm}[1]{\left\lVert#1\right\rVert}
\newcommand{\ad}[1]{\text{ad}_{S_{#1}(t)}}

\title{Topology-Induced Spontaneous Non-reciprocal Pumping in Cold-Atom Systems with Loss}


\author{Linhu Li}
\email{phylli@nus.edu.sg}
\affiliation{Department of Physics, National University of Singapore, Singapore 117551, Republic of Singapore}
\author{Ching Hua Lee} \email{calvin-lee@ihpc.a-star.edu.sg}
\affiliation{Institute of High Performance Computing, A*STAR, Singapore, 138632.}
\affiliation{Department of Physics, National University of Singapore, Singapore 117551, Republic of Singapore}
\author{Jiangbin Gong}  \email{phygj@nus.edu.sg}
\affiliation{Department of Physics, National University of Singapore, Singapore 117551, Republic of Singapore}


\date{\today}
\begin{abstract}
We propose a realistic \GJ{cold-atom quantum setting} where \GJ{nontrivial energy-band topology}
induces non-reciprocal pumping.  This is an intriguing non-Hermitian phenomenon that illustrates how topology,  when assisted with atom loss,  can act as a ``switch'' \GJ{for non-Hermitian skin effect (NHSE), rather than as a passive property
that is modified by the NHSE}.  In particular, we present a lattice-shaking scenario to realize a two-dimensional cold-atom platform, where non-reciprocity is switched on only in the presence of both atom loss and topological localization due to time-reversal symmetry breaking. \GJ{Such spontaneous non-reciprocal pumping is manifested by asymmetric dynamical evolution, detectable by atomic populations along the system edges.}  Our results may trigger possible applications in \GJ{non-reciprocal} atomtronics, where loss and topological mechanisms conspire to control atomic transport.

\end{abstract}

\maketitle



\noindent{\it Introduction. --} Cold atoms on optical lattices provide a highly promising platform for demonstrating interesting topological and many-body physics~\cite{bloch2008many,goldman2016topological}. Recent advances in lattice shaking technology have enabled unprecedented tuning of effective tunneling amplitudes, leading to pioneering observations of various exotic topological states~\cite{lignier2007dynamical,struck2011quantum,struck2012tunable,hauke2012non,jotzu2014experimental,Jin2018coldatom,Jin2019Creutz}.
Beyond realizing conventional static and Floquet topological phases~\cite{goldman2010realistic,liu2013detecting,liu2013floquet,liu2014realization,potirniche2017floquet,li2018realistic,lee2018floquet,Zhou2019floquet,yang2018dynamical},
cold-atoms in optical {systems} are also suitable \GJ{quantum} platforms for simulating higher-order topological phases~{\cite{benalcazar2017HOTI,Benalcazar2017HOTI2,schindler2018HOTI,wang2018higher,zeng2019Majorana,Luo2019HOnonH}} and non-Hermitian effects {\cite{Bender1998nonH,Rotter2009non,Jiaming2019gainloss}}, two classes of phenomena of intense current interest. Indeed, non-Hermiticity can be experimentally implemented through optically-induced depopulating losses~\cite{Jiaming2019gainloss}, and be fine-tuned to exhibit non-Hermitian topological degeneracies~\cite{xu2017weyl,Luo2019HOnonH} like exceptional points~\cite{berry2004EP,rotter2009EP,heiss2012EP,Hassan2017EP,Hu2017EP}, lines~\cite{xu2017weyl,carlstrom2018EL,moors2019ER,Wang2019ER,Yang2019EL,carlstrom2019EL,yoshida2019ER,Luo2018ER,Yoshida2019ER_2} and surfaces~\cite{zhou2019exceptional,okugawa2019exceptional}, all possessing rich geometric structure without Hermitian analogs~\cite{shen2018topological,Yin2018nonHermitian,Li2019geometric}.

\begin{figure}
\includegraphics[width=1\linewidth]{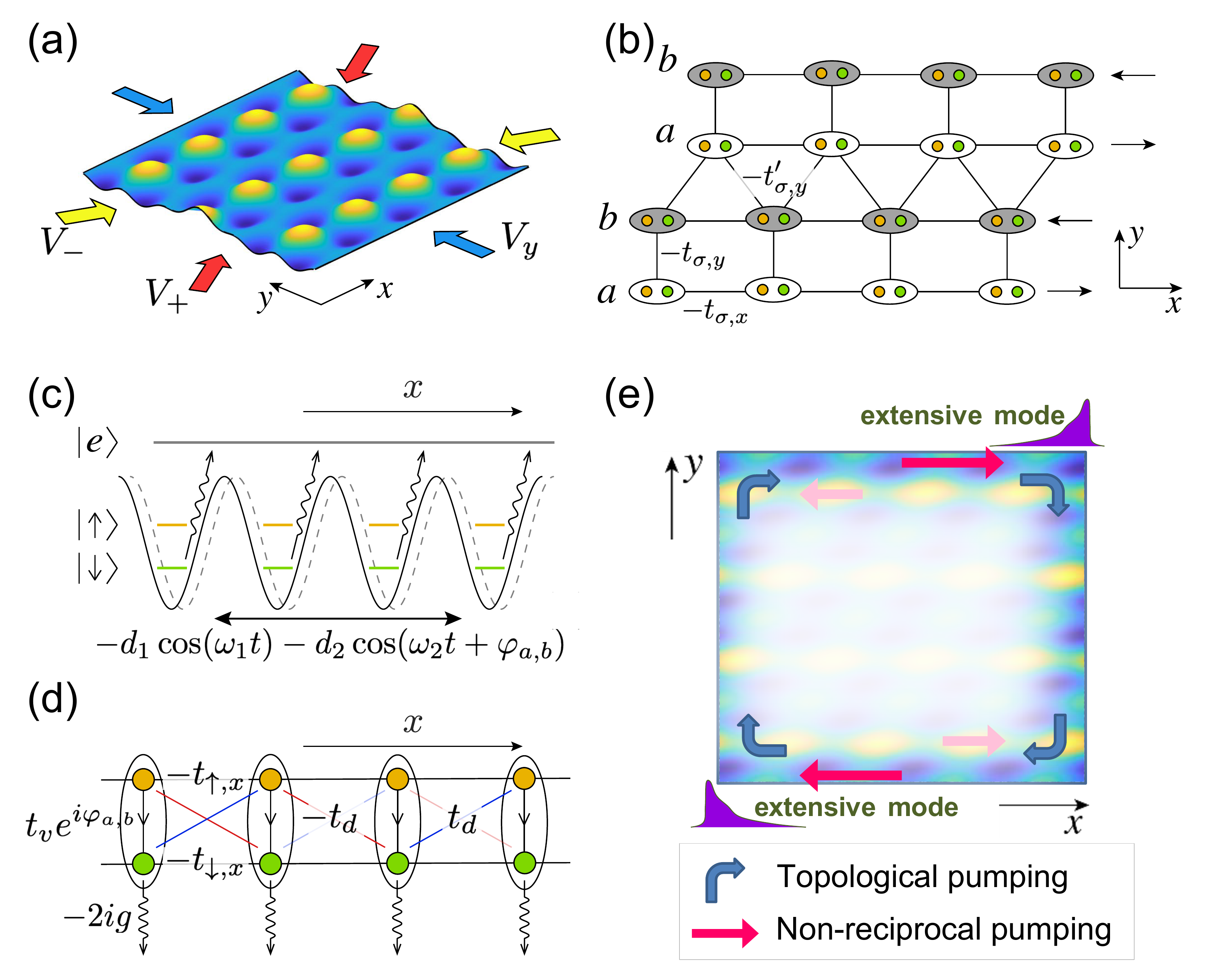}
\caption{(a) 2D optical lattice depicted by Eq.~\ref{optical_potential}, with snapshot taken when $x_1(\omega_1t)=0$ and $V_-(\omega_2 t)=V_+(\omega_2 t)=V_y/2$. (b) The corresponding lattice structure of (a), with its two sublattices shaken in opposite directions (black arrows).
(c) Effective 1D potential experienced by each sublattice along the $x$ direction, with $|\uparrow\rangle $ and $|\downarrow\rangle$ pseudospins representing the $s$ and $p_x$ orbitals. Atoms in the $|\downarrow\rangle$ state are resonantly excited to a third state $|e\rangle$, leading to loss.
(d) Two-leg system (Eq.~\ref{H2d_2}) for each sublattice (white/gray rows in Fig.~\ref{fig1:lattice}(b)), with yellow/green sites representing up/down pseudospins.
(e) Mechanism for topology-induced non-reciprocal corner modes. The reciprocity of the lattice is spontaneously broken when topological boundary modes localize on the upper/lower sublattices along the upper/lower edge, such that non-reciprocal pumping towards the right/left dominates.
}
\label{fig1:lattice}
\end{figure}

In this work, we propose the cold-atom realization of a novel phenomenon where topological localization \GJ{in one direction} spontaneously breaks the reciprocity of a two-dimensional (2D) lattice system \GJ{in the presence of atom loss}, leading to non-reciprocal pumping in the transverse direction. This is orthogonal in spirit to much contemporary theoretical~\cite{yao2018edge,kunst2018biorthogonal,xiong2018does,Song2019BBC,Lee2019anatomy,lee2018tidal,song2019non,longhi2019probing,longhi2019topological} and experimental~\cite{tobias2019observation,tobias2019reciprocal,ananya2019observation,lei2019observation} literature, which focuses on the \emph{breakdown} of conventional topological bulk-boundary correspondences (BBCs) caused by non-reciprocal pumping i.e. the non-Hermitian skin effect (NHSE). Our proposal instead illustrates how nontrivial topology can \GJ{conversely} \emph{cause} non-reciprocal pumping, as illustrated in Fig.~\ref{fig1:lattice}(e). The resultant topology-induced NHSE pumping, \GJ{with non-Hermiticity implemented via atom loss only}, is marked by corner mode accumulation scaling extensively with the system length~\cite{Lee2019hybrid}, fundamentally unlike higher-order topological corner modes, Hermitian or otherwise~\cite{Nori2019HOnonH,Edvardsson2019HOnonH,Ezawa2018HOnonH,liu2018topological,Luo2019HOnonH}.
 \GJ{Our
results may stimulate further work on non-reciprocal atomtronics, exploiting both atom loss and topological
mechanisms to control atomic transport.}
 \GJ{Relying on a \emph{quantum} platform, our proposal is also intrinsically poised for further exploring non-Hermitian effects with tunable many-body interactions.}

Our setup consists of an optically shaken lattice \GJ{accommodating cold atoms, designed such that lattice anisotropy and antiphase shaking conspire to yield nontrivial first-order topology generating one-dimensional (1D) topological edge modes.  The required atom loss is introduced through selective depopulation, by exciting the atoms into an excited state with a resonant beam.  Due to the extreme robustness of both Chern topology and the resultant nonreciprocal pumping, our topology-induced corner modes} are stable across a large region in the \GJ{parameter space}, as evident in Fig.~\ref{fig3:ST}(e). They can be distinguished from higher-order topological corner modes via their asymmetric dynamical evolution arising from tspontaneously broken reciprocity (Fig.~\ref{fig4:dynamic}).


\noindent{\it 2D optical lattice and model Hamiltonian. --} 
Consider a gas of Fermionic atoms in a two-frequency periodically driven 2D superlattice formed by three directed optical standing waves, described by the potential [Fig.~\ref{fig1:lattice}(a)]
\begin{eqnarray}
&& V(x,y)=2V_y\left[\sin^2 \left(\frac{2\pi}{\lambda_L}y\right)+\cos^2 \left(\frac{4\pi}{\lambda_L}y\right)\right]\nonumber\\
&+&2V_-(\omega_2t)
\cos^2\{ \frac{2\pi}{\lambda_L}[\beta (x-x_1(\omega_1t))+y/2]+\frac{\pi}{4}\}\nonumber\\
&+&  2V_+(\omega_2t) \sin^2\{ \frac{2\pi}{\lambda_L}[\beta (x-x_1(\omega_1t))-y/2]+\frac{\pi}{4}\},
\label{optical_potential}
\end{eqnarray}
where $\lambda_L/2$ sets the lattice constant along the $y$-direction. 
$V_y$ describes a static double-well potential along the $y$-direction, realizable with two pairs of interfering laser beams~\cite{peil2003patterned,folling2007direct,trotzky2008time,double_well_atala}. The other two potentials with coefficients $V_\pm$ have oblique orientations determined by $\beta$, and are time-dependent with oscillations controlled by frequency $\omega_1$ through $x_1(\omega_1t)=-d_1\cos\omega_1t$, induced by sinusoidally modulating the frequency difference between the interfering laser beams~\cite{gemelke2005parametric,lignier2007dynamical,Zheng2014floquet,Zhang2014shaping,Jin2019Creutz}; and small-amplitude driving with frequency $\omega_2$ through $V_{\pm}(\omega_2 t)=V_{xy}[1\pm A\cos(\omega_2 t+\varphi)]/2$, $|A|\leqslant1$.
These dynamical potential modulations of dissimilar frequencies break TR symmetry, and can induce Chern topology. In particular, the oppositely modulated $V_\pm$ amplitudes gives rise to a lattice of potential minima $x_0^\tau(t)$ with its two sublattices $\tau=a,b$ (white, gray) shaken in opposite directions [arrows in Fig.~\ref{fig1:lattice}(b)]: $x^\tau_0(t)=x_1(t)-d_2\cos(\omega_2t+\varphi_{\tau})$,
where $\varphi_a=\varphi_b+\pi=\varphi$ and effective shaking amplitude $d_2\approx\frac{d}{2\beta\pi}\tan^{-1}\left[\sqrt{5/3}A\right]$~\cite{SuppMat}.
 While our setup is valid for generic parameter values, we shall set $\beta=1$, $\lambda_L=532$ nm in the numerics that follow, and consider small shaking amplitudes $d_2=d_1/20=\lambda_L/200$~\footnote{Larger amplitudes lead to a greater number of nontrivial couplings without necessarily introducing new physics.}.

To obtain a suitable effective tight-binding Hamiltonian, we consider (unless otherwise stated) \GJ{bi-chromatic} driving~\cite{Jin2019Creutz} with resonant frequency values $\omega_2 =2\omega_1\approx \epsilon_{sp_x}/\hbar$, $\epsilon_{sp_x}$ the on-site energy difference between the $s$ and $p_x$ orbitals of the fermions. These orbitals are in turn coupled by two-photon inter-orbital resonant couplings~\cite{Zheng2014floquet,Zhang2014shaping,Jin2019Creutz} arising from a momentum term induced by $x$-direction shaking, which do not couple different $p_y$ orbitals. As such, $p_y$ orbitals are decoupled from the $p_x$ and $s$ orbitals, labeled as pseudospins $|\uparrow\rangle$ and $|\downarrow\rangle$ respectively. Finally, to introduce nontrivial non-Hermiticity, on-site atom loss is asymmetrically introduced on the $|\downarrow\rangle$ state by using a resonant optical beam to transfer the atoms 
to an excited state $|e\rangle$ [Fig. \ref{fig1:lattice}(c)]~\cite{Jiaming2019gainloss}.

In the high frequency regime with small oscillation amplitudes, the Magnus expansion approximation gives the effective \emph{static} Hamiltonian $H({\bm k})=\Psi^{\dagger}({\bm k})h_{\rm 2D}({\bm k})\Psi({\bm k})$ with $\Psi^{\dagger}({\bm k})=(\hat{a}^{\dagger}_{\uparrow{\bm k}},\hat{b}^{\dagger}_{\uparrow{\bm k}},\hat{a}^{\dagger}_{\downarrow{\bm k}},\hat{b}^{\dagger}_{\downarrow{\bm k}})$ creating states on $a,b$ sublattices and $|\uparrow\rangle,|\downarrow\rangle$ pseudospins, and~\cite{SuppMat}

\begin{eqnarray}
h_{\rm 2D}(\bm k)&=&h^+_{\sigma}(\bm k)\sigma_0+h^-_{\sigma}(\bm k)\sigma_3+h^+_{\tau}(\bm k)\tau_0+h^-_{\tau}(\bm k)\tau_3,\nonumber\\
\label{H2d_1}
\end{eqnarray}
\begin{widetext}
\begin{eqnarray}
h^{\pm}_{\sigma}(\bm k)&=&-(2t_{\pm,x}\cos k_x-\Delta_{\pm}\pm ig)\tau_0
-\{t_{\pm,y}+t'_{\pm,y}[\cos k_y +\cos (k_y-k_x)]\} \tau_1
-t'_{\pm,y}[\sin k_y +\sin (k_y-k_x)] \tau_2,\nonumber\\
h^{-}_{\tau}(\bm k)&=&(t_v\cos\varphi)\sigma_1+(t_v\sin\varphi)\sigma_2,~h^{+}_{\tau}(\bm k)=(2t_d\sin k_x)\sigma_2,\label{H2d_2}
\end{eqnarray}
\end{widetext}
\noindent $\tau_i$ and $\sigma_i$ ($i=1,2,3$) two sets of Pauli matrices acting on the sublattice and pseudospin spaces respectively, with $\tau_0$ and $\sigma_0$ their corresponding two by two identity matrices. The various coupling amplitudes $t_v,t_d$, $t_{\pm,\alpha}=(t_{\uparrow,\alpha}\pm t_{\downarrow,\alpha})/2$ and $t'_{\pm,\alpha}=(t'_{\uparrow,\alpha}\pm t'_{\downarrow,\alpha})/2$ arise from overlap integrals between the atomic orbitals, as detailed in the Supplementary Material~\cite{SuppMat}. Since these couplings do not connect different pseudospin and sublattice components simultaneously, $h_{\rm 2D}(\bm k)$ can be visualized as the lattice in Fig.~\ref{fig1:lattice}(b) with each $\tau\in \{a,b\}$ row represented by a two-leg lattice with couplings between $s$ and $p_x$ orbitals differing by a $\pi$ phase i.e. $\varphi_a=\varphi_b+\pi$ ( Fig.~\ref{fig1:lattice}(d)). The shaking-induced inter-orbital coupling $t_v$ and $t_d$ breaks the TR and inversion symmetries respectively, giving rise to Chern topology~\cite{jotzu2014experimental}.   \GJ{Note that,
 non-Hermiticity enters here through $-ig\tau_0 (\sigma_0-\sigma_3)$, depicting a loss mechanism acting only on the $|\downarrow\rangle$ pseudospin sector. This pseuedospin down sector is energetically offset from the $|\uparrow\rangle$ sector by $\Delta_-$.   There is no atom gain introduced to the system.}  The  $\Delta_+$ term represents an overall energy shift that we can be neglected).
For concreteness in our numerics, we have considered a lattice loaded with a Fermionic gas of $^{173}$Yb atoms, although our scheme is also applicable for other cold atoms.

\begin{figure}
\includegraphics[width=1\linewidth]{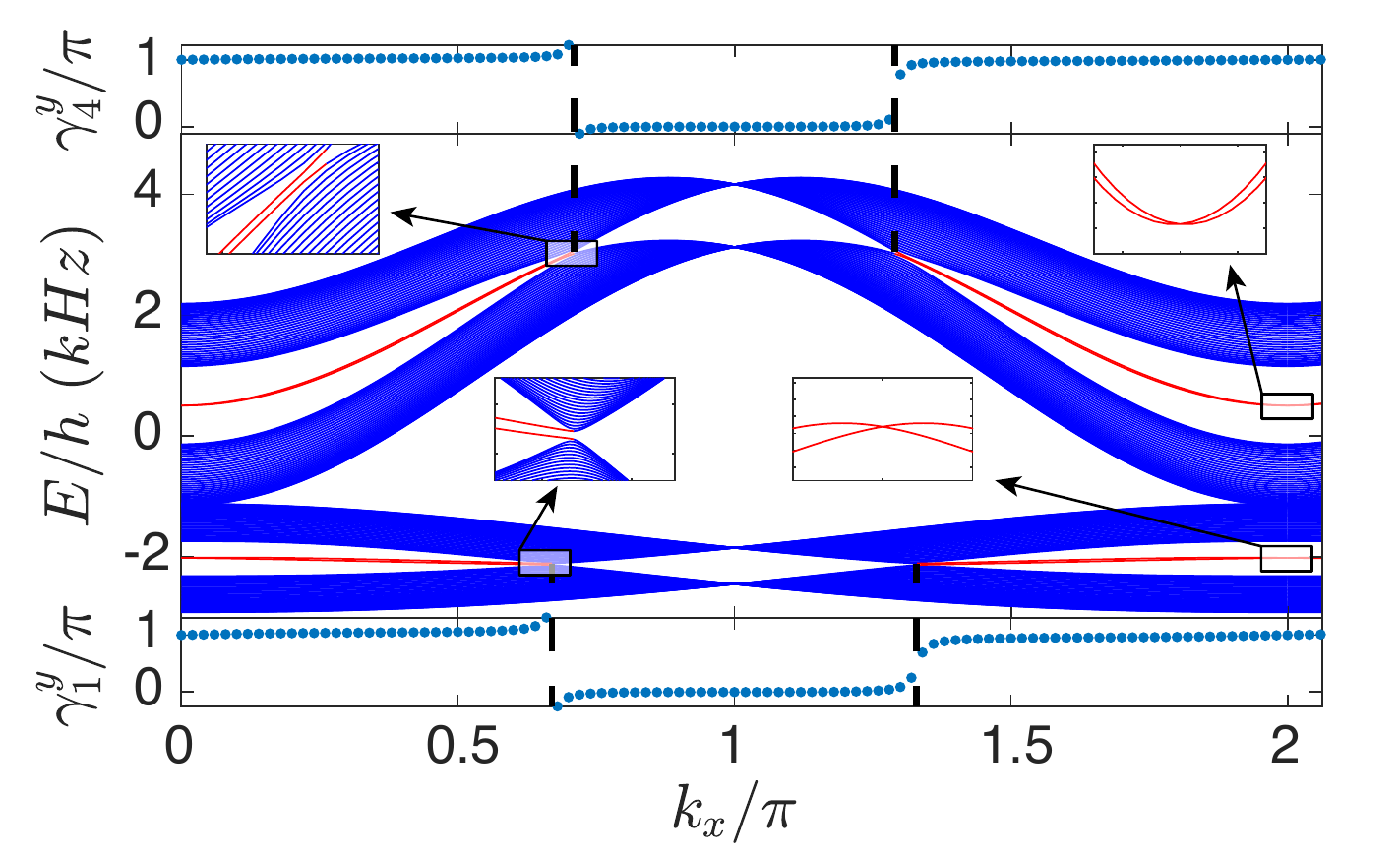}
\caption{$x$-PBC/$y$-OBC spectrum of $h_{\rm 2D}$ without atom loss. Almost degenerate chiral edge modes (red curves) reflect the Chern topology, with their presence/absence at each $k_x$ depending on whether the nearly-quantized Berry phases $\gamma_1^y,\gamma_4^y$ are close to $\pi$ or $0$.
Effective parameters are $\varphi=\pi/2$ and $\{t_{+,x},t_{-,x},t_{+,y},t_{-,y},t'_{+,y},t'_{-,y},\Delta_-,t_d,t_v\}=\{0.38,0.43,-0.11,-0.41,0.44,0.14,1.20,-0.11,0.19\}$ $h\times$kHz, corresponding to the experimentally realistic parameters $V_{xy}=3E_r=\frac{3}{2}V_y$ and $\omega_2=2\omega_1=10\times2\pi$ kHz, with $E_r=\frac{\pi^2\hbar^2}{2md^2}=h \times 4.1$ kHz.
}
\label{fig2:topology}
\end{figure}

\begin{figure}
\includegraphics[width=1\linewidth]{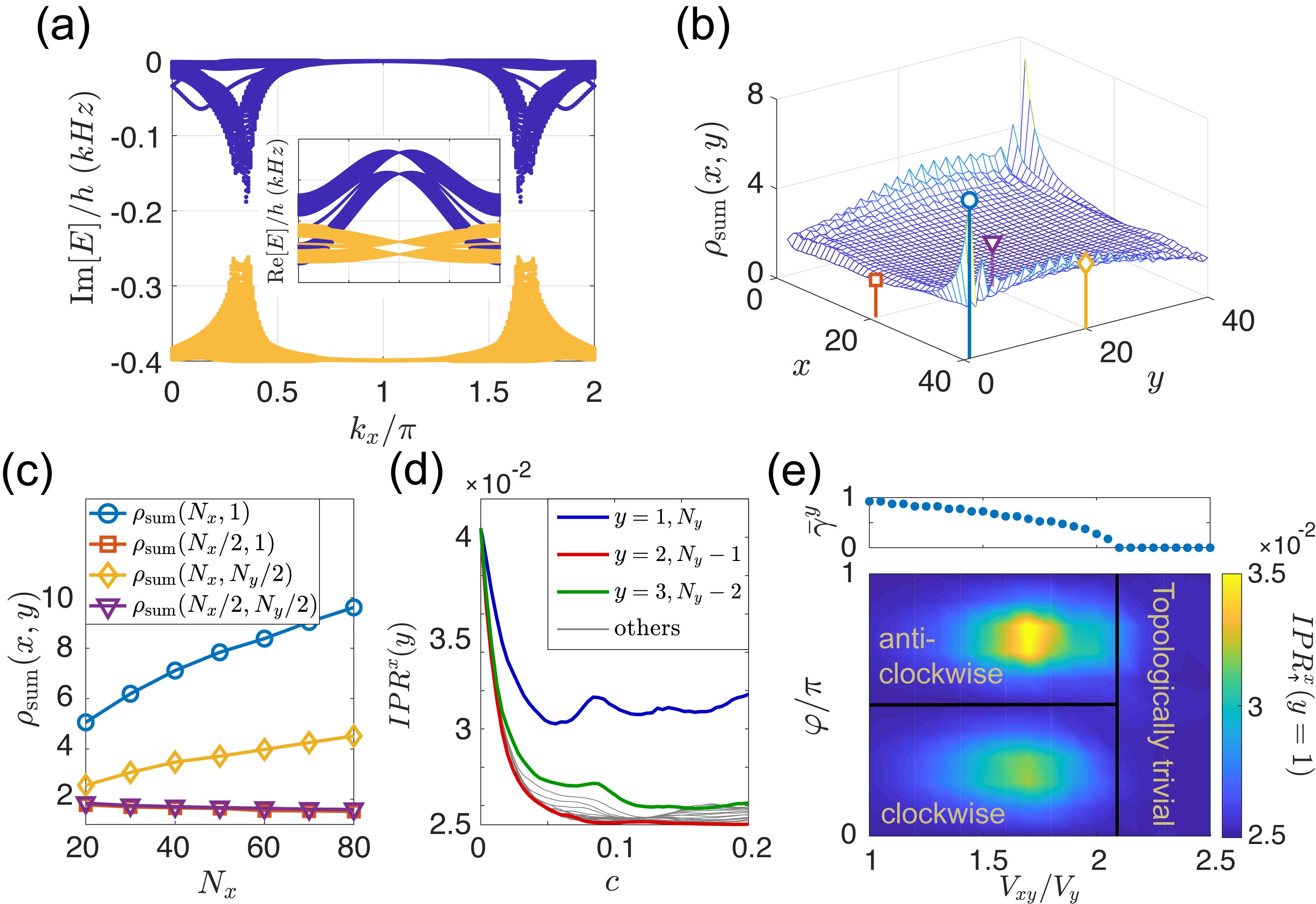}
\caption{(a) Imaginary and real parts of the $x$-PBC/$y$-OBC spectrum for nonzero loss with $g=0.2$ $h\times$kHz, color indicating different bands.
(b) The corresponding summed eigenmode distribution $\rho_{\rm sum}(x,y)$, with corner mode densities $> \mathcal{O}(1)$, much greater than those of ordinary topological corner modes.
(c) Scaling of $\rho_{\rm sum}(x,y)$ with system length $N_x$ at various positions indicated in (b), with the NHSE giving rise to linear i.e. extensive scaling at the corner $(N_x,1)$ (blue), and to a smaller extent along an edge $(N_x,N_y/2)$ (yellow), but not deep in the bulk at $(N_x/2,N_x/2)$ (purple) and the other edge at (Nx/2,1) (red).
(d) The $x$ direction IPR for different sublattices $y$ as a function of inter-sublattice coupling extent $c$, computed for $N_x=40$. While higher IPR$^x$ (boundary accumulation) occurs at all $y$ in the decoupled ($c=0$) limit, only the top/bottom-most sublattices still experience IPR$^x > 1/N_x$ i.e. boundary (corner) accumulation when $c$ increases.
(e) Phase diagram showing anti/clockwise topology-induced nonreciprocal pumping phases over a large range of optical lattice parameters $\phi$ and $V_{xy}/V_y\gtrsim2.1$. The average Berry phase $\bar{\gamma}^y$ along the top edge $y=1$ is nonzero in the topological phases, where the IPR$^x_\uparrow$ for pseudospin-up component also peaks. The shaking frequency is set as $2\omega_1=\epsilon_{sp_x}/\hbar$. 
}
\label{fig3:ST}
\end{figure}

\begin{figure}
\includegraphics[width=1\linewidth]{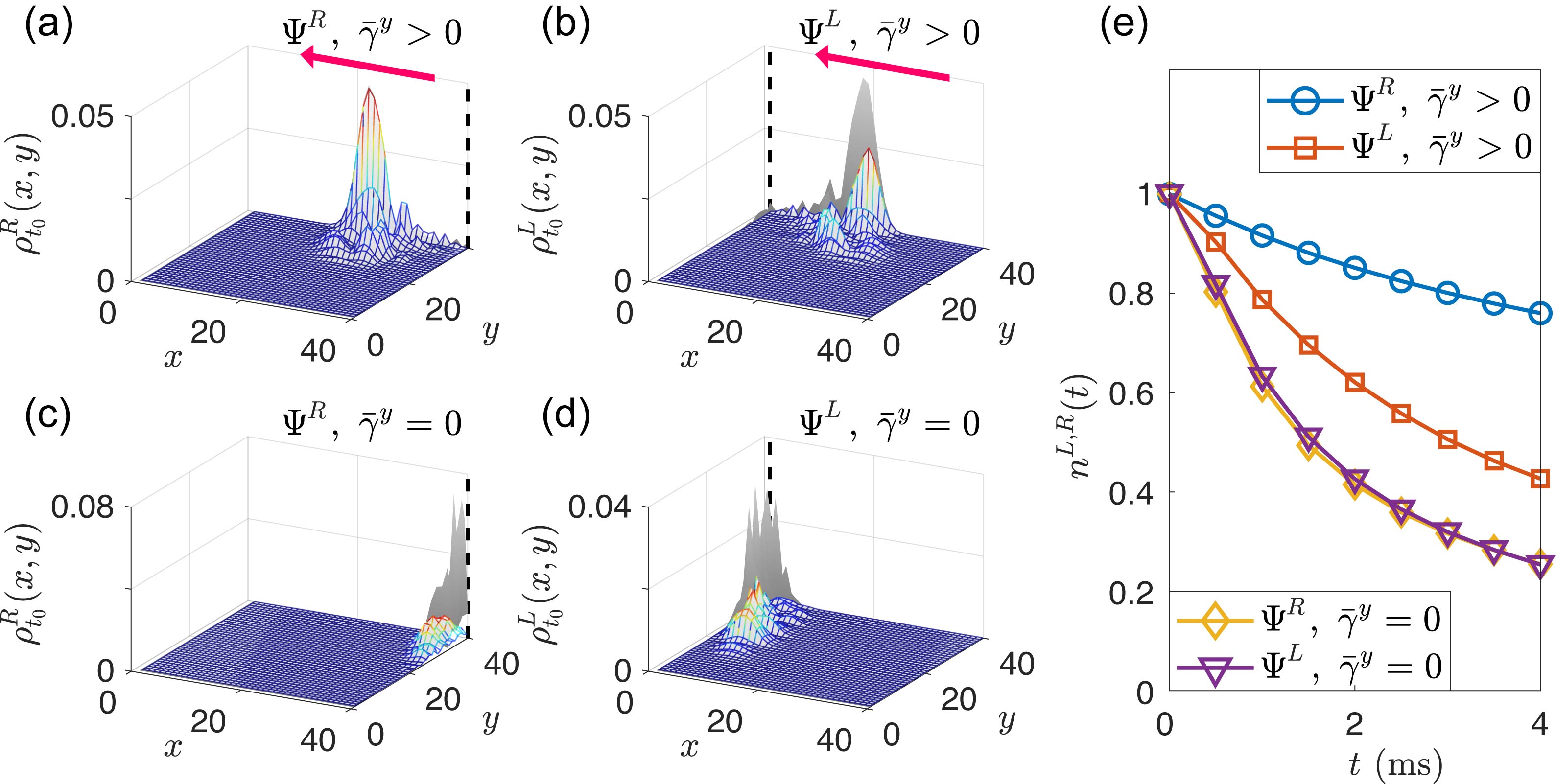}
\caption{(a-d) Population distribution snapshots of the states $\Psi^{L/R}_t$ after $t_0=2$ ms of evolution with $g=0.5~h\times$kHz, as experimental signature of topology-induced  corner accumulation. The initial state is prepared at $(\sigma,x_0,y_0)=(\uparrow,1,N_y)$ in (a,c), and $(\uparrow,N_x,N_y)$ in (b,d), as indicated by the dashed lines. Pink arrows in (a,b) indicate the direction of corner accumulation, with $\Psi^R$ state spreading in the same direction suffering less decay. For comparison, gray shadows indicate the same states evolving without dissipation. Other parameters are $\varphi=\pi/2$, $V_y=2E_r$, and (a,b) $V_{xy}=3E_r$, $\omega_1=6$ kHz; (c,d) $V_{xy}=4.5E_r$, $\omega_1=9$ kHz. In (a), $\Psi^R$ evolves almost identically as the no-loss case and hence covering the gray shadow.
(e) Evolution of total population $n^{L/R}$ for cases (a-d) with initial state launched from different corners  and topology. Case (a) i.e. $\Psi^R$ with $\bar{\gamma}^y>0$ (blue circles) decays least, manifesting topology-induced non-reciprocal pumping.
}
\label{fig4:dynamic}
\end{figure}

\noindent{\it Edge modes from topological localization. --}
In the static ($\omega_1=\omega_2=0$) Hermitian limit, our $h_{\rm 2D}(\bm k)$ system posseses sublattice symmetry, and is topologically nontrivial with $Z$-quantized 1D Berry phase
\begin{eqnarray}
\gamma^y_n(k_x)=-{\rm Im}\oint_0^{2\pi} dk_y \langle u_n(k_x,k_y)|\partial_{k_y} | u_n(k_x,k_y)\rangle
\end{eqnarray}
for each $n$-th band $|u_n(k_x,k_y)\rangle$. When TR symmetry is broken through lattice shaking at nonzero $\omega_2=2\omega_1$~\cite{SuppMat}, its resultant topological boundary modes along the $y$-edge crossover to Chern chiral edge modes, analogous to the zigzag-edge edge modes of Graphene or other honeycomb lattices under a circularly polarized laser~\cite{usaj2014irradiated,perez2014floquet,sentef2015theory}. As evident in Fig. \ref{fig2:topology} with open boundary conditions (OBCs) along $y$, these Chern edge states (red) are however only weakly separated due to the small shaking amplitudes and hence small inter-pseudospin coupling. This weak coupling also yields \emph{almost quantized} Berry phases $\gamma_1^y$ and $\gamma_4^y$ shown in  Fig.~\ref{fig2:topology}.
The average Berry phase $\bar{\gamma}^y_n=\sum_{k_x}\gamma^y_n(k_x)/N_x$ used later gives the ratio of the total number of 1D edge modes to $N_x$ under $x$-PBC/$y$-OBC, which turns out to be proportional to the corner mode accumulation strength \cite{Lee2019hybrid} and hence spontaneous non-reciprocal pumping elaborated below.

\noindent{\it Topology-induced nonreciprocal corner modes.--} The non-Hermiticity introduced by on-site atom loss does not merely perturb the spectrum to complex values [Fig. \ref{fig3:ST}(a)].
Interestingly, it also produces strong mode accumulation at two opposite corners, as conceptually sketched in Fig.~\ref{fig1:lattice}(e) and quantitatively plotted in Fig.~\ref{fig3:ST}(b). Fundamentally unlike corner modes arising from higher-order topology, our corner mode density scales \emph{extensively} with the system length, as elucidated in Fig.~\ref{fig3:ST}(c) by the linear scaling behavior of the total mode intensity $\rho_\text{sum}(x,y)=\sum_{n,\sigma}|\psi_{n}(\sigma,x,y)|^2$ with system length $N_x$ at the nontrivial corner $(x,y)=(N_x,1)$. (or at the other corner $(1,N_y)$, not shown). Since the number of topological boundary modes is fixed by the topological invariant, this extensive scaling of mode intensity must have alternatively originated from some form of the NHSE \footnote{The NHSE is characterized by extensive mode accumulation along a boundary in the simultaneous presence of non-reciprocity and non-Hermiticity.}. \GJ{Indeed, we have realized one form of hybrid corner modes \cite{Lee2019hybrid} by exploiting atom loss.}

To understand this enigmatic corner mode accumulation, first note that \GJ{the loss term $-ig\tau_0 (\sigma_0-\sigma_3)$} in the $|\downarrow\rangle$ sector hardly changes the topological modes, as evident from comparing the Re$[E]$ inset of Fig.~\ref{fig3:ST}(a) with the dissipationless bandstructure in Fig.~\ref{fig2:topology}.
With chosen experimental parameters producing considerably weak inter-orbital couplings $t_v$ and $t_d$, one can also largely confine the damping effects within the $|\downarrow\rangle$ sector (orange in Fig.~\ref{fig3:ST}(a) with large -Im$[E]$), and observe non-reciprocal effects in the $|\uparrow\rangle$ sector possessing almost real eigenenergies.

We can obtain intuition about how these non-reciprocal effects can be topology-induced from a Gedanken experiment of tuning the couplings $(t_{\sigma,y},t'_{\sigma,y})\rightarrow c(t_{\sigma,y},t'_{\sigma,y})$, where $c\in[0,1]$ interpolates between the decoupled limit and the full values of these couplings. In the fictitious $c=0$ decoupled limit, $t_{\sigma,y}=t'_{\sigma,y}=0$, and the system is decoupled into a series of 1D two-leg ladders as in \GJ{Fig.~\ref{fig1:lattice}(d)}, with $\varphi_{a,b}$ differing by $\pi$ between the sublattices $a$ and $b$.
\GJ{Upon rotating $\sigma_1$, $\sigma_2$ and $\sigma_3$, the decoupled 1D two-leg ladders of each sublattice resemble the Su-Schrieffer-Heeger (SSH) model with non-Hermitian hoppings \cite{SuppMat}. Hence they exhibit extensive skin mode accumulation at either the left ($x=1$) or right ($x=N_x$) edge.  This is already an interesting aspect of our optical lattice design insofar as our system is reciprocal. The combined motion of both sublattices is still reciprocal when $\sin\varphi_a=-\sin\varphi_b$ \cite{SuppMat}, which is automatically satisfied since $\varphi_a-\varphi_b=\pi$. Nevertheless,
topological localization on one sublattice breaks this delicate reciprocity.}

The \emph{physical} spatial eigenmode distribution is obtained by restoring the $t_{\sigma,y}$ and $t'_{\sigma,y}$ couplings, i.e. by tuning $c$ from $0$ to $1$. This couples the two SSH-like sublattices,  leading to destructive interference of the equal and opposite NHSEs unless both sublattices are not symmetrically occupied. Since the $\tau=a,b$ sub-lattices of our $h_{\rm 2D}(\bm k)$ take similar forms, bulk modes are necessarily equitably distributed, and thus experience very little residual non-reciprocal pumping and hence mode accumulation. However, its topological modes along the top ($y=N_y$) and bottom ($y=1$) are intrinsically sublattice-polarized.  As such, we expect the top-most and bottom-most rows of sites to experience equal and opposite uncancelled NHSEs, leading to extensive mode accumulation at two of the four corners (Fig.~\ref{fig1:lattice}(f)). This mode accumulation within each sublattice row $y$ can be \GJ{quantified} by the inverse participation ratio (IPR) in the $x$ direction:
\begin{eqnarray}
\text{IPR}^x(y)= \sum_{x}\left[\frac{\sum_{\sigma,n}|\psi_{n}(\sigma,x,y)|^2}{\sum_x\sum_{\sigma,n}|\psi_{n}(\sigma,x,y)|^2}\right]^2
\end{eqnarray}
with ${\psi}_{n}(\sigma,x,y)$ the $n$-th eigenstate of the system, as plotted for various rows $y$ as a function of $c$ in Fig. \ref{fig3:ST}(d).
In particular, in the fictitious decoupled limit $c=0$, $\text{IPR}^x(y)$ takes the same value for different $y$, indicative of equally strong NHSE for each decoupled 1D two-leg ladder. As $c$ increases to physically realistic values, the ladders couple, and $\text{IPR}^x(y)$ rapidly decrease to the uniform limit $1/N_x$ in the bulk ($1<y<N_y$), implying full delocalization along the $x$ direction. However, at the bottom/top boundaries at $y=1,N_y$ (blue curves), $\text{IPR}^x(y)$ stabilizes at values much higher than $1/N_x$ as $c$ is increased, thereby indicative of left/right mode accumulation which, in 2D, gives corner mode accumulation [Fig.~\ref{fig3:ST}(b)].  \GJ{The obtained non-reciprocal pumping can be also regarded as a strong manifestation of topological localization along $y$.}



These observations are further digested in the phase diagram of Fig.~\ref{fig3:ST}(e). Intuitively, corner accumulation should be expected if 1D edge states exist in the non-dissipative limit, e.g., whenever the average Berry phase $\bar{\gamma}^y\equiv\bar{\gamma}^y_4$ of the least dissipative (4-th) band is non-vanishing. In addition, Fig.~\ref{fig3:ST}(e) reveals that the corner mode localization, now quantified by the IPR$^x_{\uparrow}$ \GJ{of the $|\uparrow\rangle$ sector only \CH{(since the least dissipative band is almost $|\uparrow\rangle$-polarized)}}, also varies strongly with optical lattice parameters $\varphi$ and $V_{xy}/V_y$. Indeed, a weaker $V_{xy}$ indicates stronger Hermitian couplings along the $x$ direction, which effectively weakens the nonreciprocal pumping. 
Also, the NHSE completely disappears when $\varphi=0$ or $\pi$ restores the TR symmetry. Since $\bar{\gamma}^y=0$ when $V_{xy}/V_y\gtrsim 2.1$ (i.e., no more edge states), most pronounced corner accumulation occurs at moderate $V_{xy}/V_y$ and $\varphi=\pm \pi/2$, with  the sign \GJ{ determining the chirality of the corner accumulation}.

\noindent{\it Dynamical behavior from non-reciprocal pumping. --} While the IPR can measure the extent of induced localization, its purported non-reciprocal origin has to be verified through dynamical behavior. We compare the time evolution of initial states $\Psi_{\rm ini}^{L/R}=\sum_{x,y}\sum_{\sigma=\uparrow,\downarrow}\psi_{\rm ini}^{L/R}(\sigma,x,y)\hat{\tau}^\dagger_{\sigma,x,y}|0\rangle$ localized on the left/right corners, i.e. $\psi_{\rm ini}^L(\uparrow,1,N_y)=1$ and $\psi_{\rm ini}^R(\uparrow,N_x,N_y)=1$ respectively, with vanishing amplitudes for $|\downarrow\rangle$ and all other sites. For these two cases, the evolved states at time $t$ are given by
$\Psi_{t}^{L/R}=e^{-i H t/\hbar}\Psi_{\rm ini}^{L/R}.$
The presence of topology-induced non-reciprocal pumping can be saliently observed in snapshots of $\Psi_{t}^{L/R}$ after a fixed time $t=t_0$. Figs.~\ref{fig4:dynamic}(a-d) illustrate the spatial densities $\rho_{t_0}^{L/R}(x,y)=\sum_\sigma|\psi_{t_0}^{L/R}(\sigma,x,y)|^2$, $\sigma =\uparrow,\downarrow$ for a representative $t_0=2$ ms, with (a,c) showcasing $\rho_{t_0}^R$ and (b,d) showcasing $\rho_{t_0}^L$, as indicated by the vertical dashed lines. In the presence of topological edge modes (i.e. nonzero $\bar{\gamma}^y$), the eigenstates are asymmetrically distributed across the sublattices, and are thus subjected to asymmetric gain/loss as it spreads in either direction. As illustrated in Figs.~\ref{fig4:dynamic}(a,b) for the case with right-to-left non-reciprocal pumping at $y=N_y$ row, $\Psi_t^L$ [Fig.~\ref{fig4:dynamic}(b)] spreads from the left corner towards the right, and is significantly more attenuated than $\Psi_t^R$ [Fig.~\ref{fig4:dynamic}(a)] that is launched from the left.  On the other hand, when topological localization is absent [Figs.~\ref{fig4:dynamic}(c,d)], both $\Psi_t^R$ and $\Psi_t^L$ dissipate rapidly with time, and do not even propagate much along the boundaries \GJ{albeit having the same loss mechanism.}
This asymmetry in dynamical propagation can be further quantified through the decaying of the population
\begin{eqnarray}
n^{L/R}(t)=\sum_{x,y}\rho_{t}^{L/R}=\sum_{x,y}\sum_\sigma|\Psi_{t}^{L/R}(\sigma,x,y)|^2.
\end{eqnarray}
In Fig.~\ref{fig4:dynamic}(e), the initial state $\Psi^R$ in the nontrivial ($\bar\gamma^y>0$) phase, depicted by Fig.~\ref{fig4:dynamic}(a), indeed gets to decay the slowest. Both $\Psi^R,\Psi^L$ in the non-topological cases [Figs.~\ref{fig4:dynamic}(c,d)] decays even faster than $\Psi^L$ with $\bar\gamma^y>0$.  

\noindent{\it Discussion. --}\GJ{With most non-Hermitian effects realized only in {\it classical} systems, simulating non-Hermitian physics
in  {tunable} {\it quantum} systems is still in its infancy.
 Via lattice shaking, we have proposed a versatile quantum platform accommodating both topological band structures and atom loss. In opposite conceptual flow to existing literature focusing on non-reciprocal pumping as a route towards unconventional topological BBCs, our proposal instead shows how nontrivial topology, assisted by loss, can \emph{induce} non-reciprocal pumping.   }

\GJ{Our large parameter space may afford further exploration into various topical issues.  In principle, the effective two-leg ladders of each sublattice possess 1D $Z$-type topology, and hence also support topological edge localization along the $x$ direction. The combination of such localizations along both $x$ and $y$ directions can lead to higher-order topological corner modes~\cite{li2018direct} in an extended parameter range~\cite{SuppMat}. Our proposed system also paves the way for exploring more exotic physics, such as the interplay of many-body interaction, atom loss, and topological phase transitions.}

\begin{acknowledgements}  \emph{Acknowledgements.}-- J.G. acknowledges support from the Singapore NRF grant No.~NRF-NRFI2017-04 (WBS No.~R-144-000-378-281).
\end{acknowledgements}

\bibliography{references}

\clearpage

\onecolumngrid
\begin{center}
\textbf{\large Supplementary Materials}\end{center}
\setcounter{equation}{0}
\setcounter{figure}{0}
\renewcommand{\theequation}{S\arabic{equation}}
\renewcommand{\thefigure}{S\arabic{figure}}

\section{Derivation of the 2D tight-binding model}
\subsection{The lattice potential}
\begin{figure}
\includegraphics[width=1\linewidth]{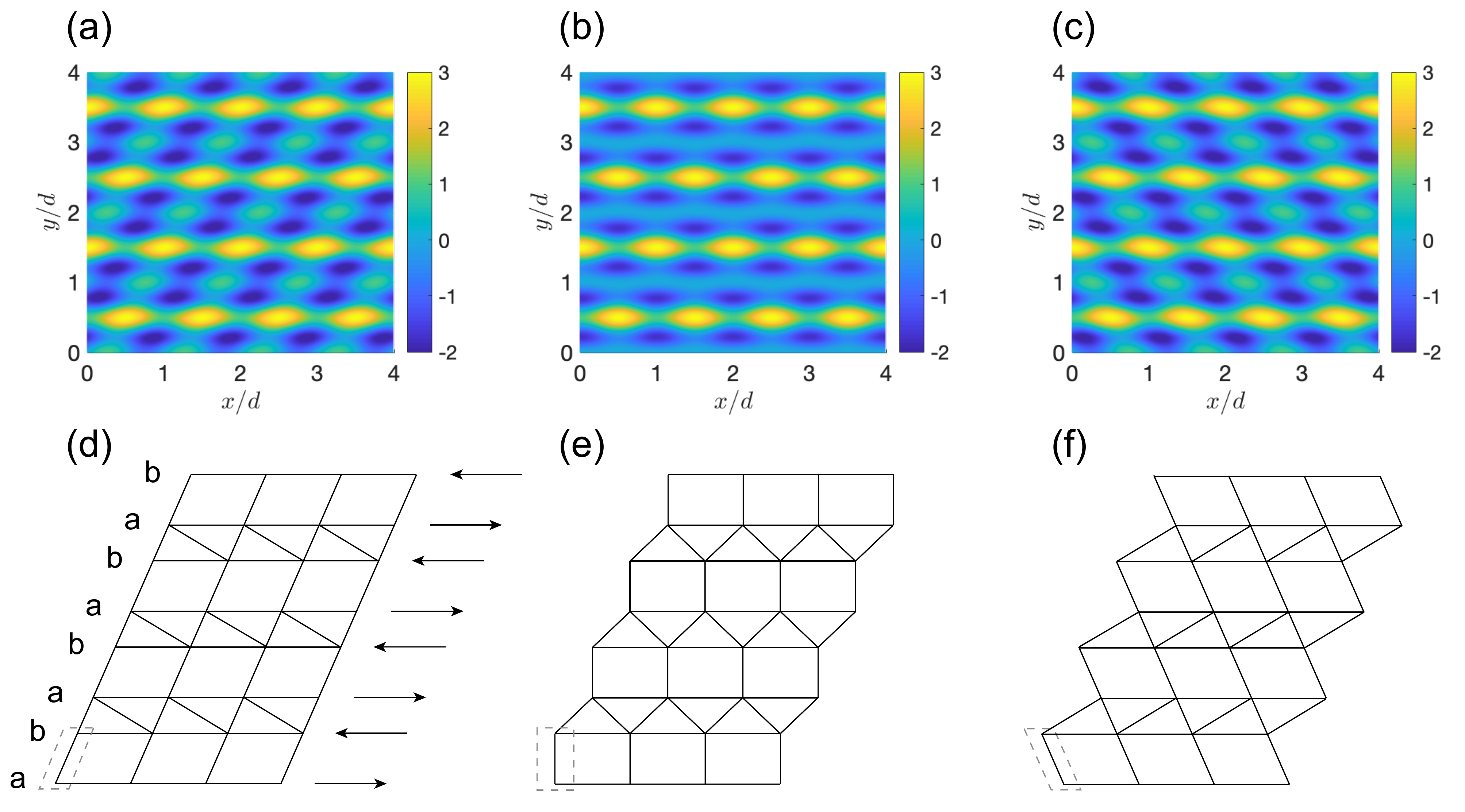}
\caption{(a-c) The spatial profile of the optical potential of $V(x,y)$, with $x_1=0$, $V_y=1$, and $(V_+,V_-)=(1,0)$, $(0.5,0.5)$, and $(0,1)$ from (a) to (c) respectively. (d-f) Sketches of the lattice structure corresponding to panels (a-c).}
\label{optical_lattice}
\end{figure}

Our cold-atom setup is described by a 2D superlattice formed by three sets of optical standing waves, described by the lattice potential
\begin{eqnarray}
V(x,y)&=&V_y\left[2\sin^2 \left(\frac{\pi}{d}y\right)+2\cos^2 \left(\frac{2\pi}{d}y\right)\right]\nonumber\\
&&+2V_{-}(\omega_2 t)
\cos^2\{ \frac{\pi}{d}[\beta (x-x_1(\omega_1 t))+y/2]+\frac{\pi}{4}\}\nonumber\\
&&+  2V_{+}(\omega_2 t) \sin^2\{ \frac{\pi}{d}[\beta (x-x_1(\omega_1 t))-y/2]+\frac{\pi}{4}\},\nonumber\\
&=&V_y(-\cos[\frac{2\pi}{d}y]+\cos[\frac{4\pi}{d}y])+2V_y\nonumber\\
&&-V_{-}(\omega_2 t) \sin\{ \frac{2\pi}{d}[\beta (x-x_1(\omega_1 t))+y/2]\}\nonumber\\
&&+V_{+}(\omega_2 t) \sin\{ \frac{2\pi}{d}[\beta (x-x_1(\omega_1 t))-y/2]\}+V_{xy},\nonumber\\
\label{S_potential}
\end{eqnarray}
where $x_1(t)$ represents a uniform lattice position shaking along $x$ direction,
\begin{eqnarray}
x_1(\omega_1 t)=-d_1\cos(\omega_1 t),
\end{eqnarray}
which can be induced by sinusoidally modulating the frequency difference between the two interfered laser beams~\cite{gemelke2005parametric,lignier2007dynamical,Zheng2014floquet,Zhang2014shaping,Jin2019Creutz}.
{The double-well potential $V_y$ induces two sublattices in our system, allowing the emergence of 1D nontrivial topology and edge modes along $y$-direction. The lattice constant along $y$ direction is given by $d=\lambda_L/2$ with $\lambda_L$ the wave-length of the corresponding laser beams. The other two potentials $V_{\pm}$ have oblique orientations determined by $\beta$, which are mirror-symmetric to each other regarding the $y$-axis. Therefore, by tuning off and on these two oblique potentials alternatively (through the $\omega_2$ driving), we can have the lattice structure switching between different configurations, and the positions of the two sublattices shifting toward opposite directions, as shown in Fig. \ref{optical_lattice}.
Here we choose
\begin{eqnarray}
V_{\pm}=\frac{1\pm f(\omega_2 t)}{2},
\end{eqnarray}
thus the switching between the two oblique potentials can be conveniently described by the varying of $f(\omega_2 t)$ from $-1$ to $1$. The explicit form of $f(\omega_2 t)$ will be defined in latter discussion when analyzing the shaking of the two sublattices.
We also note that while in the main text we have set $\beta=1$, the following discussion is for a generic value of $\beta$, unless specified otherwise.}

\subsection{Positions of the lattice sites}
We first derive the time-dependence of the effective lattice site positions by solving for the potential minima. The partial derivatives of the potential $V(x,y)$ give
\begin{eqnarray}
\frac{\partial V(x,y)}{\partial x}&=&\frac{2\pi \beta}{d} V_{xy}\left[f(\omega_2 t)\cos\frac{2\pi\beta(x -x_1)}{d}\cos\frac{\pi y}{d}+\sin\frac{2\pi\beta(x -x_1)}{d}\sin\frac{\pi y}{d}\right],\nonumber\\
\frac{\partial V(x,y)}{\partial y}&=&\frac{2\pi}{d}V_y(1-4\cos \frac{2\pi y}{d})\sin \frac{2\pi y}{d}-
\frac{2\pi}{d}\frac{V_{xy}}{2}\left[f(\omega_2 t)\sin\frac{2\pi\beta(x -x_1)}{d}\sin\frac{\pi y}{d}+\cos\frac{2\pi\beta(x -x_1)}{d}\cos\frac{\pi y}{d}   \right].\nonumber\\
\end{eqnarray}
At oscillation turning points $f(\omega_2 t)=\pm1$, the stationary points of the potential satisfy
\begin{eqnarray}
\pm\cos\frac{2\pi\beta(x -x_1)}{d}\cos\frac{\pi y}{d}+\sin\frac{2\pi\beta(x -x_1)}{d}\sin\frac{\pi y}{d}&=&0,\nonumber\\
 \left(1-4\cos \frac{2\pi y}{d}\right)\sin \frac{2\pi y}{d}&=&0,
\end{eqnarray}
and the minima of $V(x,y)$ are found to be at
\begin{eqnarray}
\frac{2\pi}{d}(x_0^a,y_0^a)_{n_x,n_y}&=&\left[\frac{2\pi}{d}(x_1\pm d_2^a)+(n_y-1)\pi+2(n_x-1)\pi
,\arccos\frac{1}{4}+2(n_y-1)\pi\right],\nonumber\\
\frac{2\pi}{d}(x_0^b,y_0^b)_{n_x,n_y}&=&\left[\frac{2\pi}{d}(x_1\mp d_2^b)+(n_y-1)\pi+2(n_x-1)\pi
,(2\pi-\arccos\frac{1}{4})+2(n_y-1)\pi\right],\label{omega2_pm1}
\end{eqnarray}
with
\begin{eqnarray}
d_2^a=-d_2^b=\frac{d}{4\beta\pi}(\arccos\frac{1}{4}-\pi).
\end{eqnarray}
When $f(\omega_2 t)=0$, the minima of $V(x,y)$ are found to be at
\begin{eqnarray}
\frac{2\pi}{d}(x_0^a,y_0^a)_{n_x,n_y}&=&\left[\frac{2\pi}{d}x_1+(n_y-1)\pi+2(n_x-1)\pi
,\frac{2\pi}{d}d_y^a+2(n_y-1)\pi\right],\nonumber\\
\frac{2\pi}{d}(x_0^b,y_0^b)_{n_x,n_y}&=&\left[\frac{2\pi}{d}x_1+(n_y-1)\pi+2(n_x-1)\pi
,\frac{2\pi}{d}d_y^b+2(n_y-1)\pi\right],\label{omega2_0}
\end{eqnarray}
{with $d_y^a$ and $d_y^b$ the first two minima of the potential at $x=0$ and $f(\omega_2 t)=0$,}
\begin{eqnarray}
V(x=0,y)=V_y(-\cos\left[\frac{2\pi}{d}y\right]+\cos\left[\frac{4\pi}{d}y\right])-V_{xy}\sin \frac{\pi}{d}y+2V_y+V_{xy}.
\end{eqnarray}
{Thus the $f(\omega_2 t)$ driving induces an oscillation of $y$-minima with an amplitude of $d_{y,2}=|d_y^a-\frac{d}{2\pi}\arccos\frac{1}{4}|=|d_y^b-d+\frac{d}{2\pi}\arccos\frac{1}{4}|$, and an oscillation of $x$-minima with an amplitude of $d_{x,2}=\frac{d}{4\beta\pi}(\pi-\arccos\frac{1}{4})$.}
We can see that $d_{y,2}$ has a dependence on the ratio of $V_y/V_{xy}$, and is much smaller than $d_{x,2}$ in the parameter region we consider (e.g. $d_{y,2}/d_{x,2}\approx1/7$ for $V_{xy}=1.5V_y$ and $\beta=1$ as chosen for Fig. 2 in main text), hence we shall ignore the shaking along $y$ hereafter, and denote $d_{x,2}$ as $d_2$ for simplicity.

The time-dependent position of $x_0^\tau(t)=x_1(t)+x_2^\tau(t)$ with $\tau\in\{a,b\}$ is given by $\partial V(x,y)/\partial x=0$, which yields
\begin{eqnarray}
x_0^{\tau}&=&x_1+x_2^\tau,\nonumber\\
x_2^\tau&=&\frac{d}{2\beta\pi}\left[\arctan\left[ -f(\omega_2 t)\cot\left(\frac{\pi y_0^{\tau}}{d}\right)\right]+2(n_x-1)\pi\right],\label{omega2}
\end{eqnarray}
with $y_0^\tau$ the $y$ position of the $\tau$ sublattice. As we have omitted the small shaking along $y$ direction, here we use
\begin{eqnarray}
\frac{2\pi}{d}y_0^a=\arccos\frac{1}{4}+2(n_y-1)\pi,~~\frac{2\pi}{d}y_0^b=2\pi-\arccos\frac{1}{4}+2(n_y-1)\pi,
\end{eqnarray}
i. e. the results from Eq. (\ref{omega2_pm1}).
As in the main text, we choose
\begin{eqnarray}
f(\omega_2 t)=A\cos(\omega_2 t +\varphi),
\end{eqnarray}
with $\varphi$ a phase factor,
and the position shifting of $x_2^\tau$ has a time period of $2\pi$.
From Eq. (\ref{omega2}) we can also see that $x_2^a=-x_2^b$, which can be represented by a $\pi$ phase difference of the phase factors of the two sublattices, $\varphi_a=\varphi_b+\pi$.
By choosing $\omega_2=2\omega_1$, we can realize the bichromatic tuning \cite{Jin2019Creutz} with a $\pi$ phase difference between the two sublattices,
except that the $f(\omega_2 t)$ driving corresponds to a position shift described by Eq. \ref{omega2}.
The amplitude of the $\omega_2$ oscillation can be controlled by tuning $A$ or $\beta$, resulting in a total amplitude of $d_2=\frac{d}{2\beta\pi}\arctan\left[\sqrt{5/3}A\right]$.

\begin{figure}
\includegraphics[width=1\linewidth]{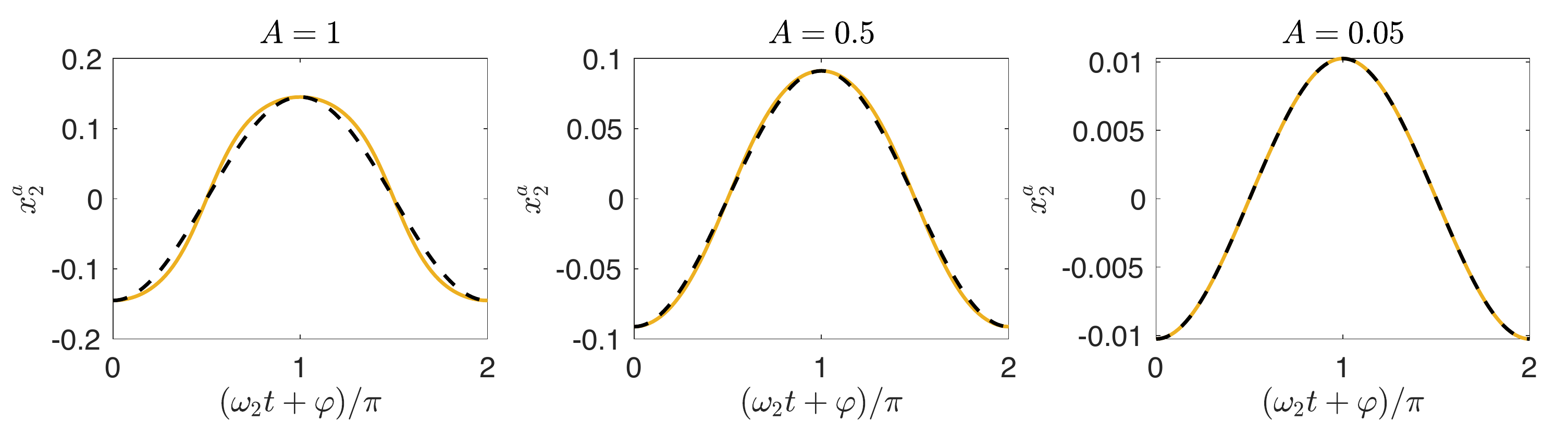}
\caption{Approximation of Eq. (\ref{omega2}) by a single fourier harmonic. The position shifting $x_2^a$ of the first unit cell with different values of $A$. The yellow solid lines are for the actual shifting obtained from Eq. (\ref{omega2}), $x_2^a=\frac{d}{2\beta\pi}\arctan\left[-\sqrt{5/3}A\cos(\omega_2 t+\varphi)\right]$, and the black dash lines are for $x_2^a=-d_2\cos(\omega_2 t+\varphi_a)$ with $d_2=\frac{d}{2\beta\pi}\arctan\left[\sqrt{5/3}A\right]$. The shaking of $b$ sublattice is given by $x_2^b=-x_2^a$. Here we choose the lattice constant to be $d=1$, and $\beta=1$.}
\label{figS2:shaking}
\end{figure}

While the actual form of the $\omega_2$ oscillation is given by Eq. (\ref{omega2}), it is very similar to a cosine function, especially when $A$ is small, as shown in Fig. \ref{figS2:shaking}.
In the main text we have considered the case with $d_2=d/100$, corresponding to $A\approx 0.05$, where the position shifting is almost identical with a cosine function.
Therefore, in the following calculations we take $x_0^{\tau}=-d_1\cos(\omega_1 t)-d_2\cos(\omega_2 t +\varphi_{\tau})$ with $\varphi_a=\varphi_b+\pi=\varphi$, so that we can apply the high frequency approximation to obtain a relatively simple tight-binding model.

\subsection{Effective Hamiltonian}
Next we derive the time-dependent Hamiltonian of the above potential. The shaking of the lattice can be effectively described by $x_0^{\tau}=-d_1\cos(\omega_1 t)-d_2\cos(\omega_2 t +\varphi_{\tau})$, and each of the two sublattices corresponds to a series of two-leg ladders with opposite shaking direction of $\omega_2$.
The stationary part of the whole 2D system can be given by the potential when each shaking component vanishes, i.e. $f(\omega_2 t)=0$ and $x_1=0$, with the lattice structure given by Fig. 1(a) in the main text. Thus the static potential is given by
\begin{eqnarray}
V_{\rm stat}(x,y)=V_y(-\cos\left[\frac{2\pi}{d}y\right]+\cos\left[\frac{4\pi}{d}y\right])-V_{xy}\cos\left[\frac{2\pi}{d}\beta x\right]\sin \frac{\pi}{d}y+2V_y+V_{xy}.\label{V_stat}
\end{eqnarray}
As each lattice site represents a minimum of the potential, we can do the Taylor expansion near it and keep up to the 2nd order term.
Due to the translational symmetry, here we consider the two lattice sites of the first unit cell $(n_x,n_y)=(1,1)$ as an example.
The static lattice position is given by
\begin{eqnarray}
(x_{0,s}^a,y_{0,s}^a)=(0,d_y^a),~~(x_{0,s}^b,y_{0,s}^b)=(0,d_y^b),
\end{eqnarray}
where $d_y^a$ and $d_y^b$ are the same as Eq. (\ref{omega2_0}).
Near these points the potential can be expanded as
\begin{eqnarray}
\tilde{V}_{\rm stat}(x,y)&=&V(x_{0,s}^\tau,y_{0,s}^\tau)+V_{xy}\frac{2\pi^2\beta^2}{d^2}\sin\left(\frac{\pi}{d}d_y^\tau\right)x^2\nonumber\\
&&+\frac{\pi^2}{d^2}\left[2V_y\cos\left(\frac{2\pi}{d}d_y^\tau\right)-8V_y\cos\left(\frac{4\pi}{d}d_y^\tau\right) +\frac{V_{xy}}{2}\sin\left(\frac{\pi}{d}d_y^\tau\right)\right](y-d_y^\tau)^2.
\end{eqnarray}
The corresponding wave-function of $H=\frac{p_x^2+p_y^2}{2m}+\tilde{V}_{\rm stat}(x,y)$ takes the same form for the two sublattices, i.e.
\begin{eqnarray}
\psi_{l_x,l_y}=\sqrt{\frac{1}{2^{l_x+l_y}l_x!l_y!}}\left(\frac{m}{\pi\hbar}\right)^{1/2}\left(\omega_x\omega_y\right)^{1/4}
\exp\left[-\frac{m}{2\hbar}(\omega_x \delta_x^2+\omega_y \delta_y^2)\right]H_{l_x}\left(\sqrt{\frac{m\omega_x}{\hbar}}\delta_x\right)H_{l_y}\left(\sqrt{\frac{m\omega_y}{\hbar}}\delta_y\right),
\end{eqnarray}
with $\delta_x=x-x_{0,s}^{\tau}$, $\delta_y=y-y_{0,s}^{\tau}$,
$l_{x/y}$ being the quantum numbers of the harmonic oscillators along $x$ and $y$ directions,
\begin{eqnarray}
H_l(z)=(-1)^le^{z^2}\frac{d^n}{dz^n}\left(e^{-z^2}\right)
\end{eqnarray}
the Hermite polynomials, and
\begin{eqnarray}
\omega_x&=&\sqrt{\frac{V_{xy}}{m}\sin\left(\frac{\pi}{d}d_y^\tau\right)}\frac{2\pi\beta}{d},\nonumber\\
\omega_y&=&\frac{2\pi}{d}\sqrt{\frac{1}{m}\left[V_y\cos\left(\frac{2\pi}{d}d_y^\tau\right)-4V_y\cos\left(\frac{4\pi}{d}d_y^\tau\right) +\frac{V_{xy}}{4}\sin\left(\frac{\pi}{d}d_y^\tau\right)\right]}
\end{eqnarray}
the effective frequencies of the oscillators, which take the same values for the two sublattices. The corresponding energy levels are given by
\begin{eqnarray}
E_{l_x,l_y}=\hbar\left[\omega_x(l_x+\frac{1}{2})+\omega_y(l_y+\frac{1}{2})\right]+V(x_{0,s}^\tau,y_{0,s}^\tau).
\end{eqnarray}
This wave-function approximately gives the Wannier state of the system.
Here the $s$ orbital is given by $(l_x,l_y)=(0,0)$, and there are two $p$ orbitals of $p_x:~(l_x,l_y)=(1,0)$ and $p_y:~(l_x,l_y)=(0,1)$. However, as shown later, the inter-orbital couplings are given solely by the periodic shaking, which involves only the first derivative of $x$. Therefore $p_y$ orbital does not couple with the rest two as they involve different $y$ orbitals, and we shall only consider the subspace of $s$ and $p_x$ orbital in the following discussion. For simplicity, we shall refer to the $p_x$ orbital as $p$ orbital hereafter.
Note that in the main text we have labeled the $p$ and $s$ orbitals as pseudopin up and down respectively, to give a simpler picture of the system. In this supplementary material, however, we shall use the language of $p$ and $s$ orbitals, to have a clearer description of how the two orbitals interact with each other under the periodic shaking.


With the above wave-function, we can now calculate the coupling strengths and construct a tight-binding model of the system.
{While we consider a 2D system, the shaking of the lattice positions is mostly along $x$-direction for the parameters we choose, making the lattice structure along $y$-direction irrelevant to the shaking-induced inter-orbital couplings.}
{This is because this shaking only induces a $x$-momentum term to the Hamiltonian, which commutes with the $y$-component of the Wannier state. Since the coupling strength is given by the overlap integral of the Wannier states and the Hamiltonian, the $y$-component shall only give a coefficient of unity to this term if the Wannier states are of the same $y$-orbital at the same $y$-position, and $0$ otherwise.}
{Therefore, as the total position shaking $x_0^\tau$ has a dependence on the sublattice index $\tau\in\{a,b\}$, we can use different effective Hamiltonians $H^\tau$ for the two sublattices to determine the shaking-induced couplings,
\begin{eqnarray}
H^{\tau}=\frac{p_x^2+p_y^2}{2m}+V_y\left(-\cos\left[\frac{2\pi}{d}y\right]+\cos\left[\frac{4\pi}{d}y\right]\right)-V_{xy}\sin\left[\frac{\pi}{d}y\right]\cos\left[\frac{2\pi}{d}\beta[x-x_0^\tau(t)]\right]+2V_y+V_{xy}.
\end{eqnarray}
By transforming to the co-moving frame~\cite{Mei2014topological,Zheng2014floquet,Jin2019Creutz}, $x\rightarrow x+x_0^\tau(t)$, the Hamiltonians becomes $H^\tau=H_{\rm stat}+\delta H^\tau(t)$ with the same stationary part given by
\begin{eqnarray}
H_{\rm stat}
&=&\frac{p_x^2+p_y^2}{2m}+V_y\left(-\cos\left[\frac{2\pi}{d}y\right]+\cos\left[\frac{4\pi}{d}y\right]\right)-V_{xy}\sin\left[\frac{\pi}{d}y\right]\cos\left[\frac{2\pi}{d}\beta x\right]+2V_y+V_{xy},
\end{eqnarray}
which is consistent with the the stationary part of the original potential, given by Eq. (\ref{V_stat}).}
The time-dependent terms are given by
\begin{eqnarray}
\delta H^{\tau}(t)&=&-\omega_1d_1\sin(\omega_1 t)p_x-\omega_2 d_2\sin(\omega_2 t+\varphi_{\tau})p_x.
\end{eqnarray}
Thus the tight-binding Hamiltonian is given by
\begin{eqnarray}
H_{\rm TB}&=&\sum_{n_x,n_y}\sum_{\tau=a,b}
\{\Psi^{\dagger}_{\tau,{n_x,n_y}}K_\tau(t)\Psi_{\tau,{n_x,n_y}}-[\Psi^{\dagger}_{\tau,{n_x,n_y}}J_{x,\tau}(t)\Psi_{\tau,{n_x,n_y}+1}+H.c.]\}\nonumber\\
&&-\sum_{n_x,n_y}[\Psi^{\dagger}_{a,{n_x,n_y}}J_y(t)\Psi_{b,{n_x,n_y}}+\Psi^{\dagger}_{b,{n_x,n_y}}J'_{y,+}(t)\Psi_{a,{n_x,n_y+1}}+\Psi^{\dagger}_{b,{n_x,n_y}}J'_{y,-}(t)\Psi_{a,{n_x-1,n_y+1}}+H.c.],
\end{eqnarray}
where $\Psi^{\dagger}_{\tau,{n_x,n_y}}=(\hat{\tau}^{\dagger}_{{n_x,n_y},p},\hat{\tau}^{\dagger}_{{n_x,n_y},s})$,
$\hat{\tau}^{\dagger}_{j,\sigma}$ with $\hat{\tau}\in\{\hat{a},\hat{b}\}$ is the creation operator for the atom on lattice site $(n_x,n_y)$ in orbital $\sigma\in\{p,s\}$, $s$ and $p$ correspond to the orbitals of $(l_x,l_y)=(0,0)$ and $(1,0)$ respectively.
The $K$ and $J$ matrices $K(t)$ are given by
\begin{eqnarray}
K_\tau(t)&=&\left(\begin{matrix}
\epsilon_{p} & -i[h_{0}^{sp}\sin(\omega_1 t)+\tilde{h}_{0}^{sp}\sin(\omega_2 t+\varphi_{\tau})] \\
i[h_{0}^{sp}\sin(\omega_1 t)+\tilde{h}_{0}^{sp}\sin(\omega_2 t+\varphi_{\tau})] & \epsilon_{s} \end{matrix}
\right),
\nonumber\\
J_{x,\tau}(t)&=&\left(\begin{matrix}
t_{p}-i[h_{1}^{pp}\sin(\omega_1 t)+\tilde{h}_{1}^{pp}\sin(\omega_2 t+\varphi_{\tau})] & i[h_{1}^{sp}\sin(\omega_1 t)+\tilde{h}_{1}^{sp}\sin(\omega_2 t+\varphi_{\tau})]  \\
-i[h_{1}^{sp}\sin(\omega_1 t)+\tilde{h}_{1}^{sp}\sin(\omega_2 t+\varphi_{\tau})] & t_{s}-i[h_{1}^{ss}\sin(\omega_1 t)+\tilde{h}_{1}^{ss}\sin(\omega_2 t+\varphi_{\tau})] \end{matrix}
\right),\nonumber\\
J_y&=&\left(\begin{matrix}
t_{p}^{\rm intra} & 0  \\
0 & t_{s}^{\rm intra} \end{matrix}
\right),~~
J'_{y,\pm}=\left(\begin{matrix}
t_{p,\pm}^{\rm inter} & 0  \\
0 & t_{s,\pm}^{\rm inter}. \end{matrix}
\right),
\end{eqnarray}
with
\begin{eqnarray}
\epsilon_{\sigma}&=&\int \int dx dy \psi_{\sigma}(x,y-y_0^{\tau}) H_{\rm stat} \psi_{\sigma}(x,y-y_0^{\tau}) ,\nonumber\\
t_{\sigma}&=&-\int \int dx dy \psi_{\sigma}(x,y-y_0^{\tau}) H_{\rm stat} \psi_{\sigma}(x-\frac{d}{\beta},y-y_0^{\tau}),\nonumber\\
h_{l}^{\sigma\sigma'}&=&\hbar\omega_1 d_1
\int \int dx dy \psi_{\sigma}(x,y-y_0^{\tau}) {\partial_x} \psi_{\sigma'}(x-\frac{ld}{\beta},y-y_0^{\tau}),\nonumber\\
\tilde{h}_{l}^{\sigma\sigma'}&=&\hbar\omega_2 d_2
\int \int dx dy \psi_{\sigma}(x,y-y_0^{\tau}) {\partial_x} \psi_{\sigma'}(x-\frac{ld}{\beta},y-y_0^{\tau}),\nonumber\\
t^{\rm intra}_{\sigma}&=&-\int\int dx dy \psi_{\sigma}(x,y-y_0^a) H_{\rm stat} \psi_{\sigma}(x,y-y_0^b),\nonumber\\
t^{\rm inter}_{\sigma,\pm}&=&-\int\int dx dy \psi_{\sigma}(x,y-y_0^b) H_{{\rm stat}} \psi_{\sigma}(x\mp\frac{d}{2\beta},y-y_0^a-d).
\end{eqnarray}
Here $\phi_{\sigma}$ is given by $\psi_{l_x,l_y}$ with $(l_x,l_y)=(1,0)$ [$(0,0)$] for $\sigma=p$ ($s$), and each of the first four integrals takes the same value for $\tau=a$ or $b$.

The tight-binding Hamiltonian can be written in terms of two sets of Pauli matrices and identity matrices, $\sigma_i$ and $\tau_i$ with $i=0,1,2,3$, acting on the $sp$-orbital space and the $ab$-sublattice space respectively. Here we rewrite the Hamiltonian in momentum space, i.e.
\begin{eqnarray}
H_{\rm TB}(\bm k)=\sum_{\bm k}
\Psi^{\dagger}_{\bm k}h_{\rm TB}(\bm k) \Psi_{\bm k}
\end{eqnarray}
with $\Psi^{\dagger}_{\bm k}=(\hat{a}_{p,{\bm k}},\hat{a}_{s,{\bm k}},\hat{b}_{p,{\bm k}},\hat{b}_{s,{\bm k}})$, and
\begin{eqnarray}
h_{\rm TB}(k_x,k_y)&=&
\left[\bar{\epsilon}-2t_{+,x}\cos k_x-2h_{1,+}^x\sin k_x \sin (\omega_1 t)\right]\sigma_0\tau_0-2\tilde{h}_{1,+}^x\sin k_x\sin(\omega_2 t+\varphi)\sigma_0\tau_3\nonumber\\
&&+\left[\frac{\epsilon_{sp}}{2}-2t_{-,x}\cos k_x-2h_{1,-}^x\sin k_x \sin (\omega_1 t)\right]\sigma_3\tau_0-2\tilde{h}_{1,-}^x\sin k_x\sin(\omega_2 t+\varphi)\sigma_3\tau_3\nonumber\\
&&+\left[h_0^{sp}\sin(\omega_1 t)+2h_1^{sp}\cos k_x\sin(\omega_1 t)\right]\sigma_2\tau_0+\left[\tilde{h}_0^{sp}\sin(\omega_2 t+\varphi)+2\tilde{h}_1^{sp}\cos k_x\sin(\omega_2 t+\varphi)\right]\sigma_2\tau_3\nonumber\\
&&-\{t_{+,y}+t'_{+,y}\left[\cos k_y+\cos(k_y-k_x)\right]\}\sigma_0\tau_1
-t'_{+,y}\left[\sin k_y+\sin(k_y-k_x)\right]\sigma_0\tau_2\nonumber\\
&&-\{t_{-,y}+t'_{-,y}\left[\cos k_y+\cos(k_y-k_x)\right]\}\sigma_3\tau_1
-t'_{-,y}\left[\sin k_y+\sin(k_y-k_x)\right]\sigma_3\tau_2,
\end{eqnarray}
with
\begin{eqnarray}
&&\bar{\epsilon}=\frac{\epsilon_p+\epsilon_s}{2},~~\epsilon_{sp}=\epsilon_p-\epsilon_s,\nonumber\\
&&t_{\pm,x}=\frac{t_p\pm t_s}{2},~~t_{\pm,y}=\frac{t^{\rm intra}_p\pm t^{\rm intra}_s}{2},~~t'_{\pm,y}=\frac{t^{\rm inter}_p\pm t^{\rm inter}_s}{2},\nonumber\\
&&h^x_{1,\pm}=\frac{h_1^{pp}\pm h_1^{ss}}{2},~~\tilde{h}^x_{1,\pm}=\frac{\tilde{h}_1^{pp}\pm \tilde{h}_1^{ss}}{2}.\nonumber
\end{eqnarray}
Finally, following Ref. \cite{Jin2019Creutz}, we consider the effects of the photon-assisted inter-orbital resonant coupling,
and obtain the Hamiltonian in a rotating reference frame with frequency $\omega_2=2\omega_1$, taking a unitary transformation of
\begin{eqnarray}
U(t)=\cos (\frac{\omega_2}{2} t)\sigma_0\tau_0-i\sin (\frac{\omega_2}{2} t)\sigma_3\tau_0.
\end{eqnarray}
The final Hamiltonian takes the form of
\begin{eqnarray}
h'_{\rm TB}(k_x,k_y)&=&
\left[\bar{\epsilon}-2t_{+,x}\cos k_x-2h_{1,+}^x\sin k_x \sin (\omega_1 t)\right]\sigma_0\tau_0-2\tilde{h}_{1,+}^x\sin k_x\sin(\omega_2 t+\varphi)\sigma_0\tau_3\nonumber\\
&&+\left[(\epsilon_{sp}-\hbar\omega_2)/2-2t_{-,x}\cos k_x-2h_{1,-}^x\sin k_x \sin (\omega_1 t)\right]\sigma_3\tau_0-2\tilde{h}_{1,-}^x\sin k_x\sin(\omega_2 t+\varphi)\sigma_3\tau_3\nonumber\\
&&+\cos(\omega_2 t)\left[h_0^{sp}\sin(\omega_1 t)+2h_1^{sp}\cos k_x\sin(\omega_1 t)\right]\sigma_2\tau_0+\cos(\omega_2 t)\left[\tilde{h}_0^{sp}\sin(\omega_2 t+\varphi)+2\tilde{h}_1^{sp}\cos k_x\sin(\omega_2 t+\varphi)\right]\sigma_2\tau_3\nonumber\\
&&+\sin(\omega_2 t)\left[h_0^{sp}\sin(\omega_1 t)+2h_1^{sp}\cos k_x\sin(\omega_1 t)\right]\sigma_1\tau_0+\sin(\omega_2 t)\left[\tilde{h}_0^{sp}\sin(\omega_2 t+\varphi)+2\tilde{h}_1^{sp}\cos k_x\sin(\omega_2 t+\varphi)\right]\sigma_1\tau_3\nonumber\\
&&-\{t_{+,y}+t'_{+,y}\left[\cos k_y+\cos(k_y-k_x)\right]\}\sigma_0\tau_1
-t'_{+,y}\left[\sin k_y+\sin(k_y-k_x)\right]\sigma_0\tau_2\nonumber\\
&&-\{t_{-,y}+t'_{-,y}\left[\cos k_y+\cos(k_y-k_x)\right]\}\sigma_3\tau_1
-t'_{-,y}\left[\sin k_y+\sin(k_y-k_x)\right]\sigma_3\tau_2.
\end{eqnarray}

Next we consider the high frequency regime of the oscillation, where the effective Floquet Hamiltonian can be given by the Magnus expansion approximation,
\begin{eqnarray}
h_{\rm eff}=h_0+\sum_{n=1}^{\infty}\frac{[h_n,h_{-n}]}{n\hbar\omega_1},
\end{eqnarray}
where $h_n$ is the $n$th Fourier component of
$h'_{\rm TB}(\bm k)$, i.e.
\begin{eqnarray}
h'_{\rm TB}=\sum_{n}h_n e^{i n \omega_1 t}.
\end{eqnarray}
Specifically, we have
\begin{eqnarray}
h_0&=&\left[\bar{\epsilon}-2t_{+,x}\cos k_x\right]\sigma_0\tau_0
+\left[(\epsilon_{sp}-\hbar\omega_2)/2-2t_{-,x}\cos k_x\right]\sigma_3\tau_0
\nonumber\\
&&+\frac{1}{2}(\tilde{h}_0^{sp}+2\tilde{h}_1^{sp}\cos k_x)\left[\sin (\varphi)\sigma_2+\cos (\varphi)\sigma_1\right]\tau_3
\nonumber\\
&&-\{t_{+,y}+t'_{+,y}\left[\cos k_y+\cos(k_y-k_x)\right]\}\sigma_0\tau_1
-t'_{+,y}\left[\sin k_y+\sin(k_y-k_x)\right]\sigma_0\tau_2\nonumber\\
&&-\{t_{-,y}+t'_{-,y}\left[\cos k_y+\cos(k_y-k_x)\right]\}\sigma_3\tau_1
-t'_{-,y}\left[\sin k_y+\sin(k_y-k_x)\right]\sigma_3\tau_2,\nonumber\\
h_{1}&=&ih_{1,+}^x\sin k_x \sigma_0\tau_0+ih_{1,-}^x\sin k_x \sigma_3\tau_0+\frac{1}{4}(h_0^{sp}+2h_1^{sp}\cos k_x)(\sigma_1+i\sigma_2)\tau_0\nonumber\\
h_{2}&=&i\tilde{h}^x_{1,+}\sin k_xe^{i\varphi}\sigma_0\tau_3+i\tilde{h}^x_{1,-}\sin k_xe^{i\varphi}\sigma_3\tau_3\nonumber\\
h_{3}&=&-\frac{1}{4}(h_0^{sp}+2h_1^{sp}\cos k_x)(\sigma_1+i\sigma_2)\tau_0\nonumber\\
h_{4}&=&-\frac{1}{4}(\tilde{h}_0^{sp}+2\tilde{h}_1^{sp}\cos k_x)e^{i\varphi}(\sigma_1+i\sigma_2)\tau_3,
\end{eqnarray}
and $h_{-n}=h_{n}^\dagger$.
Thus the effective Hamiltonian is given by
\begin{eqnarray}
h_{\rm eff}&=&h_0-\frac{h_{1,-}^xh_0^{sp}\sin (k_x)+h_{1,-}^xh_1^{sp}\sin (2k_x)}{\hbar\omega_1}\sigma_2\tau_0\nonumber\\
&&+\frac{(h_0^{sp})^2+4h_0^{sp}h_1^{sp}\cos (k_x)+2(h_1^{sp})^2+2(h_1^{sp})^2\cos (2k_x)}{3\hbar\omega_1}\sigma_3\tau_0\nonumber\\
&&+\frac{(\tilde{h}_0^{sp})^2+4\tilde{h}_0^{sp}\tilde{h}_1^{sp}\cos (k_x)+2(\tilde{h}_1^{sp})^2+2(\tilde{h}_1^{sp})^2\cos (2k_x)}{16\hbar\omega_1}\sigma_3\tau_3.
\end{eqnarray}
With $\hbar\omega_1>h_0^{sp}\gg h_1^{sp}$, and $h_l^{\sigma\sigma'}\gg \tilde{h}_l^{\sigma\sigma'}$ (providing $d_1\gg d_2$), the Hamiltonian is approximated as
\begin{eqnarray}
h_{\rm eff}&\approx&\left[\Delta_+-2t_{+,x}\cos k_x\right]\sigma_0\tau_0
+\left[\Delta_--2(t_{-,x}-\frac{2h_0^{sp}h_1^{sp}}{3\hbar\omega_1})\cos k_x\right]\sigma_3\tau_0\nonumber\\
&&+t_v\cos k_x\left[\sin (\varphi)\sigma_2+\cos (\varphi)\sigma_1\right]\tau_3
-2t_d\sin (k_x)\sigma_2\tau_0\nonumber\\
&&-\{t_{+,y}+t'_{+,y}\left[\cos k_y+\cos(k_y-k_x)\right]\}\sigma_0\tau_1
-t'_{+,y}\left[\sin k_y+\sin(k_y-k_x)\right]\sigma_0\tau_2\nonumber\\
&&-\{t_{-,y}+t'_{-,y}\left[\cos k_y+\cos(k_y-k_x)\right]\}\sigma_3\tau_1
-t'_{-,y}\left[\sin k_y+\sin(k_y-k_x)\right]\sigma_3\tau_2,
\end{eqnarray}
with
\begin{eqnarray}
&&\Delta_+=\bar{\epsilon},~~\Delta_-=(\epsilon_{sp}-2\hbar\omega_1)/2+\frac{(h_0^{sp})^2}{3\hbar\omega_1},\nonumber\\
&&t_v=\frac{1}{2}\tilde{h}_0^{sp},~~t_d=\frac{h_{1,-}^xh_0^{sp}}{2\hbar\omega_1}.\nonumber
\end{eqnarray}
Note the here the term $\frac{2h_0^{sp}h_1^{sp}}{3\hbar\omega_1}$ is added to $t_{-,x}$, thus it also contributes to the intra-orbital couplings of $t_p$ and $t_s$ along $x$ direction. The parameters $t_{\uparrow,x}$, $t_{\downarrow,x}$, and $t_{-,x}$ in the main text are given by $t_p-\frac{2h_0^{sp}h_1^{sp}}{3\hbar\omega_1}$, $t_s+\frac{2h_0^{sp}h_1^{sp}}{3\hbar\omega_1}$, and $t_{-,x}-\frac{2h_0^{sp}h_1^{sp}}{3\hbar\omega_1}$ of the above discussion respectively.
{Finally, we note that when $V_y$ is much larger than $V_{xy}$, the couplings along $y$-direction become extremely weak, and the 2D model reduces to a series of decoupled 1D chains analogous to that of Ref. \cite{Jin2019Creutz}.}

\section{Edge states with inter-orbital couplings and atom loss}
\begin{figure}
\includegraphics[width=1\linewidth]{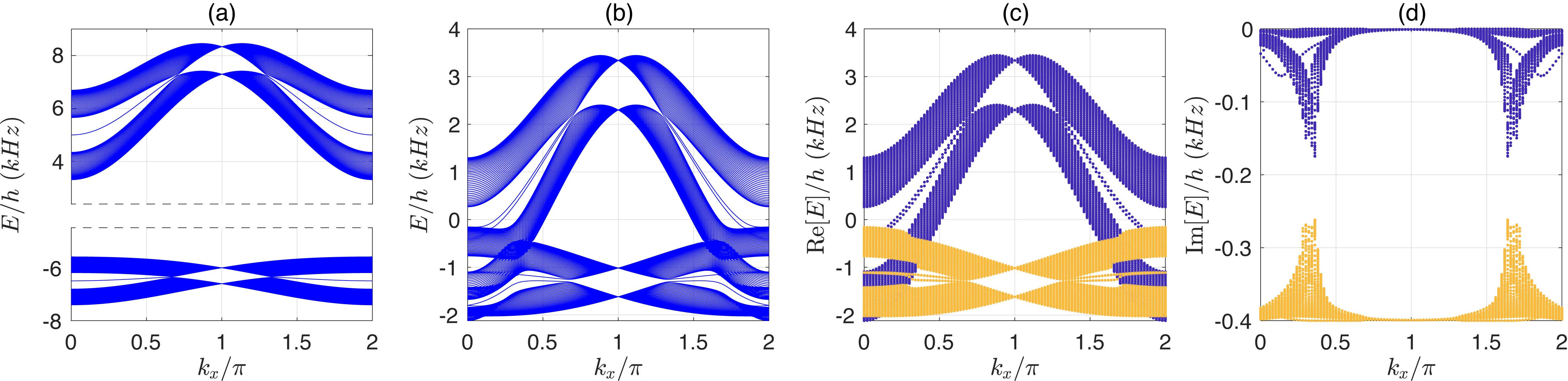}
\caption{Spectra under $x$-PBC/$y$-OBC, with (a) $\omega=0$, $g=0$, (b) $\omega=6$ kHz, $g=0$, and (c,d) $\omega=6$ kHz, $g=0.2~h\times$ kHz. The optical potential is chosen as $V_y=2E_r$ and $V_{xy}=3E_r$.}
\label{figS3}
\end{figure}

Here we elaborate on the Gedanken experiment of tuning the inter-sublattice couplings, as introduced in the main text. When the shaking frequencies $\omega_1=\omega_2=0$, the two orbitals are decoupled, resulting in two $2\times2$ subsystems described by
\begin{eqnarray}
h_p(\bm k)=-(2t_{p,x}\cos k_x-\epsilon_p)\tau_0-\{t_{p,y}+t'_{p,y}\left[\cos k_y+\cos(k_y-k_x)\right]\}\tau_1-t'_{p,y}\left[\sin k_y+\sin(k_y-k_x)\right]\tau_2
\end{eqnarray}
and
\begin{eqnarray}
h_s(\bm k)=-(2t_{s,x}\cos k_x-\epsilon_s+2ig)\tau_0-\{t_{s,y}+t'_{s,y}\left[\cos k_y+\cos(k_y-k_x)\right]\}\tau_1-t'_{s,y}\left[\sin k_y+\sin(k_y-k_x)\right]\tau_2
\end{eqnarray}
for the two orbitals respectively, each of which possesses a 1D $Z$-type topology due to the absence of the third Pauli matrix $\tau_3$.
Consequently, under $x$-PBC/$y$-OBC, the system shall have a pair of degenerate edge modes for each subsystem in a certain range of $k_x$, as shown in Fig. \ref{figS3}(a).
The high-frequency shaking we consider couples the two orbitals through the photon-assisted inter-orbital resonant couplings and mixes the energy bands of the two orbitals, as shown in Fig. \ref{figS3}(b). On the other hand, by comparing Figs. \ref{figS3}(a) and (b), we can see that the region of $k_x$ that hosts 1D edge modes remains roughly unchanged, except for some $k_x$ the edge modes overlap with the bulk bands, due to the mixing of the bands. Therefore we can still use the Berry phase for this decoupled scenario to characterize the edge modes under nonzero frequencies $\omega_1$ and $\omega_2$.

By introducing atom loss to the $s$ orbital, the two pairs of energy bands, mainly given by the two orbitals respectively, acquire different imaginary energies and thus possess an imaginary gap, as shown in Fig. (\ref{figS3})(c,d).
The edge modes are now once again fully separated from the bulk bands, while their existence is almost not affected.

\section{Non-Hermitian skin effect of the effective 1D Hamiltonian of the two-leg ladder along $x$ direction}
As discussed in the main text, when tuning off the coupling along $y$ direction, the 2D system we consider can be viewed as a series of two-leg ladders of the two sublattices $a$ and $b$.
For each sublattice $\tau=a,b$, the effective 1D Hamiltonian is $H_{\rm 1D}^{\tau,x}=\sum_{k_x} \hat{\Psi}^\dagger_{\tau,k_x}h_{\rm 1D}^{\tau,x}({k_x})\hat{\Psi}_{\tau,k_x}$, with
\begin{eqnarray}
h_{\rm 1D}^{\tau,x}({k_x})&=&-(2t_{+,x}\cos {k_x}+ig)\sigma_0+(t_v\cos\varphi_{\tau})\sigma_1\nonumber\\
&&+(t_v\sin \varphi_{\tau}+2 t_d \sin {k_x})\sigma_2\nonumber\\
&&-(2t_{-,x}\cos {k_x}-\Delta_- -ig)\sigma_3,\label{h_x_supp}
\end{eqnarray}
with $\hat{\Psi}^\dagger_{\tau,k_x}=(\hat{\tau}^\dagger_{\uparrow,k_x},\hat{\tau}^\dagger_{\downarrow,k_x})$, and $\varphi_a=\varphi_b+\pi=\varphi$.
In the simplest case of $\varphi=\pi/2$, the system possesses only two of the three Pauli matrices, and is exactly equivalent to the Su-Schrieffer-Heeger (SSH) model~\cite{SSH} upon a  basis rotation $\sigma_3\rightarrow\sigma_2\rightarrow\sigma_1\rightarrow\sigma_3$  and a $\pi/2$ shift of the quasi-momentum $k_x$. With this rotation, the non-Hermitian term of $ig\sigma_3$ in Eq.~\ref{h_x_supp} is transformed into the usual non-reciprocal coupling $ig\sigma_2$ for the SSH model, which is associated with the NHSE~\cite{yao2018edge}. Through such basis rotations, we can show that the NHSE generically occurs whenever non-Hermiticity and TR breaking are simultaneously present, even if not within the same coupling, as in dissipative on-site mechanisms on a TR-broken lattice~\cite{Fei2019nonH}.

The eigenenergies of the effective 1D Hamiltonian are given by
\begin{eqnarray}
E^{\tau}_{\pm}(k_x)=-(2t_{+,x}\cos {k_x}+ig)\pm\sqrt{f_0+f_1(k_x)+f^\tau_2(k_x)},
\end{eqnarray}
with
\begin{eqnarray}
f_0&=&t_v^2+\Delta_-^2-g^2+2i\Delta_-g,\nonumber\\
f_1(k_x)&=&4t_d^2\sin^2k_x+4t^2_{-,x}\cos^2k_x-4\Delta_-t_{-,x}\cos k_x-4it_{-,x}g\cos k_x,\nonumber\\
f_2^\tau(k_x)&=&4t_vt_d\sin\varphi_\tau\sin k_x.
\end{eqnarray}
The non-Hermitian skin effect (NHSE) in this system can be described by a non-Bloch variation of the Hamiltonian, $H(k+i\kappa_x)$, which recovers the OBC spectrum (exact for possible topological edge states), and the imaginary flux $\kappa_x$ has an correspondence to the inverse localization length of the skin localization~\cite{Lee2019anatomy,Lee2019hybrid,Li2019geometric}.
In our model, we can see that the eigenenergies satisfy
\begin{eqnarray}
E^{a}_{\pm}(k_x+i\kappa_x)=E^{b}_{\pm}(-k_x-i\kappa_x),
\end{eqnarray}
which suggests that the two-leg ladders of the two sublattices correspond to opposite skin-localization lengths, i.e. opposite directions of the nonreciprocal pumping.

\section{reflection of the non-reciprocal pumping in a finite-size system}
In a finite-size system, the non-reciprocal pumping along $x$ direction of a prepared state will eventually reach the boundary of the system. Unlike 2D topological chiral edge states which circularly move along the 1D boundaries, the motion of a wavepacket in our system is governed by a different mechanism and will be reflected when hitting the boundary. The wavepacket shall move towards the opposite direction hereafter, and hence suffers a different decaying rate.

\begin{figure}
\includegraphics[width=\linewidth]{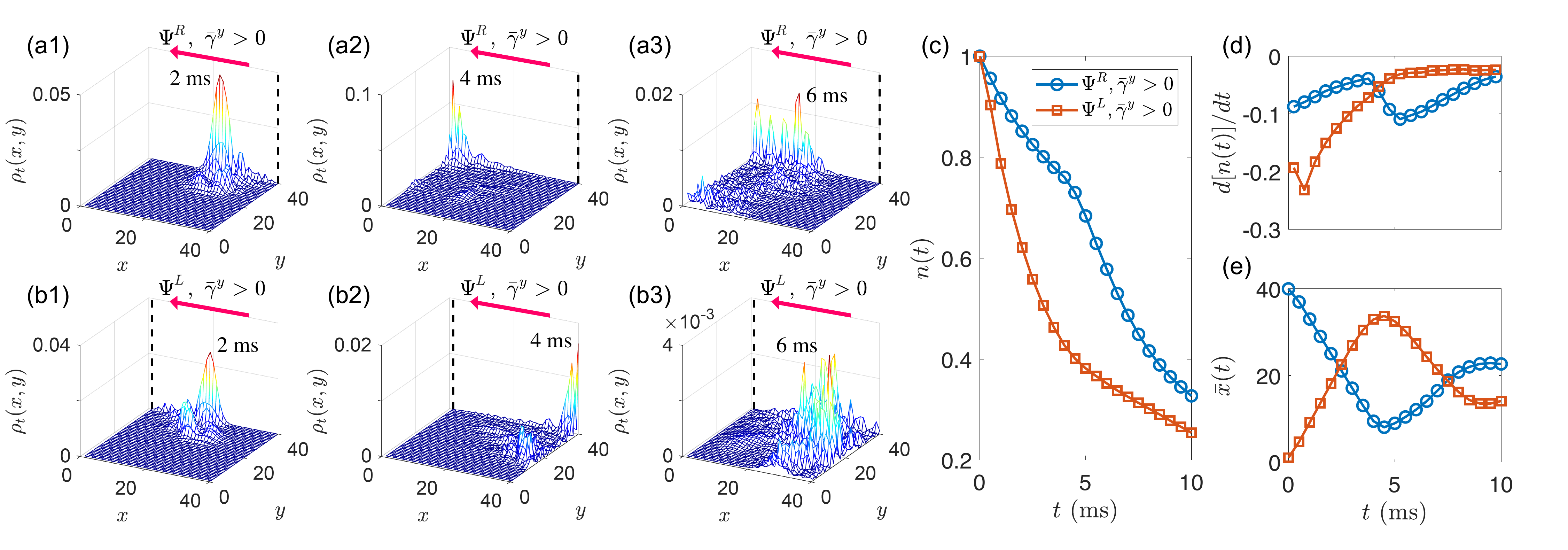}
\caption{(a,b) Population distribution of the evolved states $\Psi^{L/R}_t$ (indicated by the dash lines) at different $t$, as labeled in the figures. The pink arrows show the direction of the non-reciprocal pumping. (b) Total population as a function of $t$ the evolution time, for the two cases of (a) and (b). (c) The time derivative of the total population, and (d) The average $x$ position of the wavepacket as a function of time, corresponding to the same legend as in (c).}
\label{figS:dynamic}
\end{figure}

In Fig. \ref{figS:dynamic}(a,b), we illustrate the spatial densities at different evolution time for $\Psi^{L/R}$ defined in the main text. As we can see, in both cases the prepared states hit the other ends of the system at $\sim 4$ ms. After that, the wavepacket begins to move toward the opposite direction, which can be see from its average $x$ position shown in Fig. \ref{figS:dynamic}(e). In Fig. \ref{figS:dynamic}(c,d) we demonstrate the population distributions $n(t)$ of the two cases, and their time derivatives. We can see that
$\Psi^R$ decays faster after $\sim 4$ ms as it now moves against the non-reciprocal pumping direction [pink arrows in panel (a,b)].
On the other hand, in contrast to $\Psi_R$, the decaying rate of $\Psi^L$ does not change dramatically for the same evolution time. This is because both of $\Psi^{L/R}$ also slowly diffuse into the bulk in the evolution, and since $\Psi^L$ decays faster at the boundary, its diffusion is more dominated. As in our system the non-reciprocal pumping is only along the boundaries, a wavepacket diffused into the bulk is no longer governed by this mechanism.

\section{Second-order topological phases of the 2D lattice without dissipation.}
Here we discuss the connection between our setup and second-order topological corner modes, which also appear in a geometrically similar lattice. Such second-order modes, which are not the focus of our work, can only be mathematically realized by extending the parameters of our setup to physically not-so-realistic values.

\begin{figure}
\includegraphics[width=0.8\linewidth]{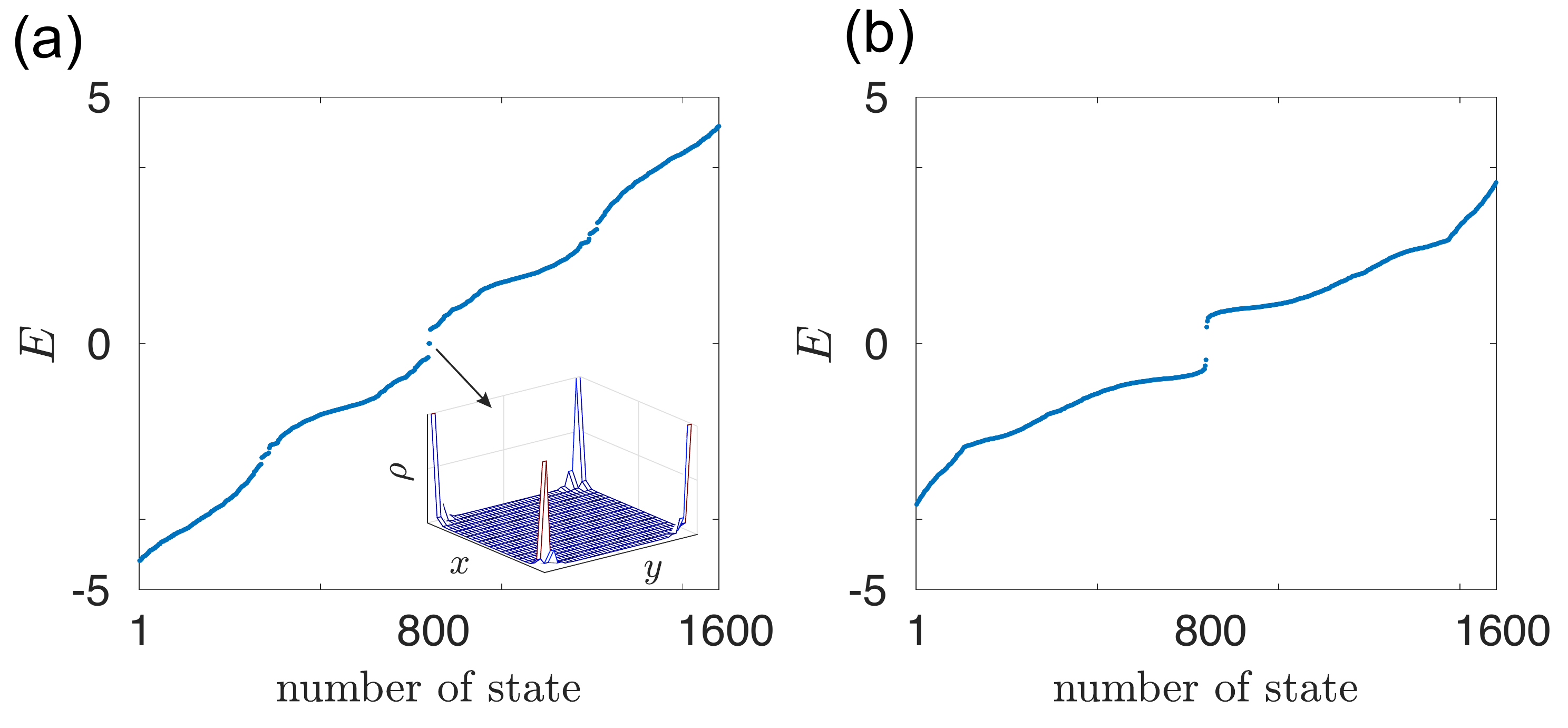}
\caption{The spectra of the system (a) with and (b) without zero-energy topological corner modes, the distribution of the corner modes are indicated in the inset of (a). The parameters are chosen as $t_{+,x} =t_{-,y} =t'_{-,y} =g=0$, $\varphi=\pi/2$, $t_{?,x} =1$, $t_{+,y} =0.5$, $t'_{+,y} = 1$, $t_v = 1$, and $t_d = 2$ and $0.3$ for (d,e) respectively.}
\label{figS:HOTI}
\end{figure}

In our construction, the effective two-leg ladders within each sublattice are analogs of the SSH model at $\varphi=\pm\pi/2$, hence they possess 1D Z-type topology and support topological edge localization also along the $x$ direction. These topological edge modes cannot induce non-reciprocal pumping along $y$ direction, as the atom loss considered here does not induce to NHSE along that direction. However, the combination of topological edge localizations along the $x$ and $y$ directions can lead to 2nd-order topological corner modes~\cite{benalcazar2017HOTI,Benalcazar2017HOTI2,li2018direct} in an extended parameter range, as shown in Fig. \ref{figS:HOTI}.
The existence of these 2nd-order topological corner modes in Fig. \ref{figS:HOTI}(a) is protected by the band gap at the zero energy. 

\end{document}